\documentclass[prd,nofootinbib,preprint,superscriptaddress]{revtex4-1}
\usepackage{amsmath}
\usepackage{graphics}
\usepackage{epstopdf}
\usepackage{graphicx}
\usepackage{subfigure}

\makeatletter
\renewcommand{\fnum@table}{\textbf{\tablename~\thetable}}
\renewcommand{\fnum@figure}{\textbf{\figurename~\thefigure}}
\makeatother

\newcommand {\be}{\begin{equation}}
\newcommand {\ee}{\end{equation}}
\newcommand {\ba}{\begin{eqnarray}}
\newcommand {\ea}{\end{eqnarray}}

\begin{document}

%%%%%%%%%%%%%%%%%%
%%  TITLE PAGE  %%
%%%%%%%%%%%%%%%%%%

\vspace*{10mm}

\title{CP-Violation and Non-Standard Interactions at the MOMENT \vspace*{1.cm} }

\author{\bf Pouya Bakhti and  Yasaman Farzan}
\affiliation{Institute for
	research in fundamental sciences (IPM), PO Box 19395-5531, Tehran,
	Iran}

\begin{abstract}
  \vspace*{.5cm}
  To measure the last unknown $3\nu$ oscillation parameter ($\delta$), several long baseline neutrino experiments have been designed or proposed. Recently it has been shown that turning on neutral current Non-Standard Interactions (NSI) of neutrinos with matter can induce degeneracies that may even hinder the proposed state-of-the-art DUNE long baseline experiment from measuring the value of $\delta$. We study how the result of the proposed MOMENT experiment with a baseline of 150 km and $200~{\rm MeV}<E_\nu<600~{\rm MeV}$ can help to solve the degeneracy induced by NSI and determine the true value of $\delta$.
\end{abstract}

\maketitle

%%%%%%%%%%%%%%%%%%%%%%%%%%%%%%%%%%%%
\section{Introduction}
%%%%%%%%%%%%%%%%%%%%%%%%%%%%
The three neutrino mass and mixing scheme has been established as the standard solution to lepton flavor violation in neutrino propagation observed by various experiments. The neutrino oscillation pattern within this scheme depends on six parameters: three mixing angles denoted by $\theta_{12}$, $\theta_{23}$ and $\theta_{13}$, a CP-violating phase $\delta$ and two mass splittings $\Delta m_{21}^2$ and $\Delta m_{31}^2$. The values of all these parameters except $\delta$ have been extracted from data. The value of $\theta_{23}$ is very close to maximal mixing value ({\it i.e.,} $\theta_{23}=45^\circ$) such that the present uncertainties do not allow to determine which octant $\theta_{23}$ belongs to. Moreover sign($\Delta m_{31}^2$) is not yet known. To determine these last unknown parameters of the neutrino oscillation scheme, an extensive experimental program is being developed. For example three setups have been suggested to determine  sign($\Delta m_{31}^2$):  i) long baseline superbeam experiments; ii) medium baseline reactor experiments, JUNO \cite{An:2015jdp} and RENO-50 \cite{Kim:2014rfa} and iii) studying the energy and zenith angle dependence of atmospheric neutrinos by giant neutrino detectors such as PINGU \cite{Aartsen:2014oha} or INO \cite{INO}. The current T2K and NO$\nu$A long baseline experiments combined with information on $\theta_{13}$ from reactor neutrino data have some limited sensitivity to the value of $\delta$. In fact, the global neutrino data analysis already shows a hint for CP-violation \cite{Thomas,Forero:2014bxa,Lisi}. According to \cite{Lisi}, at 1 $\sigma$ the allowed values of $\delta$ are in the range $\delta=(205-292)^\circ$  which includes the maximal CP-violating phase $\delta=270^\circ$ but at $3\sigma$ all values of $\delta$ are allowed. To determine the value of $\delta$ various long baseline setups have been proposed. The state-of-the-art DUNE \cite{DUNE} and T2HK \cite{T2HK} long baseline experiments
which employ conventional superbeams from pion decay will be the champions to determine the value of $\delta$. Construction of these experiments are under study. They are expected to gather enough data for determination of sign$(\Delta m_{31}^2)$ by around 2030 \cite{Patterson:2015xja}. Alternative methods to measure $\delta$ are suggested in \cite{Yas,Ge}.

There is also a proposal to build a neutrino experiment in China with a baseline of 150 km using relatively low energy ($\sim$200 MeV$- 600$ MeV).
This experiment is called MOMENT which stands for MuOn-decay MEdium baseline NeuTrino beam. The goal of MOMENT is also measuring the CP-violating phase \cite{Cao:2014bea}.
In \cite{Pilar-Moment}, the potential of MOMENT for determining $\delta$, the octant of $\theta_{23}$ and the mass ordering has been discussed and it is shown that the results of MOMENT combined with those of NO$\nu$A and T2K can help to rule out wrong solutions and dramatically  reduce uncertainties.

We emphasize that the claims mentioned above are valid only under assumption of standard interaction. New physics can give rise to new interaction of neutrinos with matter fields \cite{Yasaman1,Yasaman2} which in turn leads to modification of propagation of neutrinos in matter. In fact, the analysis of solar neutrino provides a 2 $\sigma$ hint in favor of NSI \cite{Palazzo:2011vg}. Since we cannot rule out the existence of such new physics  before experiments are carried out \cite{Pilar-DUNE,Andre-DUNE}, it is imperative to reexamine the discovery potential of these setups \cite{Poonam} in the presence of NSI. Ref. \cite{Huber} shows that the claimed preference for $\delta=270^\circ$ in the present data can be mimicked by neutrino NSI even if CP is conserved in the neutrino sector ({\it i.e.,} even if both $\delta$ and the phases of new neutrino couplings vanish). Ref. \cite{Danny} shows that although DUNE will be very efficient in solving degeneracies still some degeneracies can remain, making it impossible to determine $\delta$ in presence of NSI at 3 $\sigma$ C.L.

Both baseline ($L$) and the average neutrino energy at MOMENT are smaller than those at other long baseline experiments (T2K, NO$\nu$A and DUNE) which aim at measuring $\delta$. As a result, both standard and non-standard matter effects at MOMENT are expected to be smaller than those at T2K,
NO$\nu$A and DUNE ({\it i.e.,}  $G_F N_e\sim 0.01 |\Delta m_{31}^2|/E_\nu$ when $|\Delta m_{31}^2|L/E_\nu\sim \pi$). Thus, we expect the effects of neutral current NSI on the determination of $\delta$ by MOMENT to be small. MOMENT can therefore help to resolve this degeneracy. The aim of the present paper is to evaluate how much the results
of MOMENT can help to resolve degeneracies in determination of $\delta$ and the octant of $\theta_{23}$ in presence of neutral current NSI. Determination of mass ordering by intermediate baseline reactor experiments, JUNO and RENO-50 are not affected by
neutral current matter effects. Unless otherwise stated, we shall assume that by the time the MOMENT data release is complete, sign$(\Delta m_{31}^2)$ is already determined by JUNO and RENO-50.\footnote{There is however an exception. As shown in \cite{us}, these intermediate reactor experiment cannot distinguish between the two solutions when we simultaneously flip $\theta_{12} \leftrightarrow \pi/2 -\theta_{12}$ and $\Delta m_{31}^2 \leftrightarrow \Delta m_{23}^2$. We however dismiss this possibility for simplification.}
We also study whether MOMENT itself can determine sign($\Delta m_{31}^2$) in the presence of non-standard matter effects.

% As shown in \cite{us}, JUNO and RENO-50 suffer from a degeneracy when under $\Delta m_{31}^2\to -\Delta m_{31}^2+\Delta m_{21}^2$ and $\theta_{12}\to \pi/2-\theta_{12}$. Flipping $\theta_{12}\to \pi/2-\theta_{12}$ can be made
%compatible with solar neutrino data in presence of NSI. This is in fact the famous LMA-Dark solution \cite{upturn}                                                                                                                                                                                 which requires large lepton conserving neutral current NSI. We will check whether MOMENT can help to solve this degeneracy.

The paper is organized as follows. In sec. \ref{theo}, we review the effects of NSI on neutrino propagation in matter and the present bounds on NSI parameters. In sec \ref{char}, we review the characteristics of the MOMENT, T2K and NO$\nu$A long baseline experiments relevant for our analysis. We present our results in sec \ref{resul}. A summary is given in sec \ref{sum}.

%%%%%%%%%%%%%%%%%%%%%%%%%%%%
\section{Effects of Neutral current  NSI on neutrino oscillation\label{theo}}
%%%%%%%%%%%%%%%%%%%%%%%%%%%%%%%%%%%%%%%

The evolution of neutrino flavors in matter is governed by a Hamiltonian which  can be decomposed as follows
$$H=H_{vac}+H_{mat}$$ where in the flavor basis
$H_{vac}=U \cdot {\rm Diag} (m_1^2,m_2^2,m_3^2) \cdot U^\dagger$ and
\begin{eqnarray}
H_{mat}=\sqrt{2} G_FN_e \left( \begin{matrix} 1+\epsilon_{ee} & \epsilon_{e\mu} & \epsilon_{e\tau} \cr \epsilon_{e\mu}^* & \epsilon_{\mu\mu}
& \epsilon_{\mu\tau}\cr
\epsilon_{e\tau}^*& \epsilon_{\mu\tau}^*&\epsilon_{\tau\tau}\end{matrix}\right)
\label{Hmat}
\end{eqnarray}
where $\epsilon_{\alpha \beta}$ quantifies the effects of new physics.
 In Ref. \cite{Maltoni}, a global analysis of all neutrino oscillation data has been performed  in the presence of neutral current  NSI.
  In fact, Ref. \cite{Maltoni} presents its results in terms of $\epsilon_{\alpha \beta}^d$ and $\epsilon_{\alpha \beta}^u$ which quantify the non-standard effective four-Fermi coupling of neutrinos to the $u$ and $d$ quarks. In our notation $\epsilon_{\alpha \beta}=(N_d/N_e)\epsilon_{\alpha \beta}^d+(N_u/N_e)\epsilon_{\alpha \beta}^u$. For the earth matter, we can approximately write $N_d/N_e\simeq N_u/N_e\simeq 3$. In fitting the data, Ref \cite{Maltoni} takes $\epsilon^u$ and $\epsilon^d$ nonzero one by one. In other words,  Ref. \cite{Maltoni} finds the acceptable ranges for $\epsilon_{\alpha \beta}^u$ (for  $\epsilon_{\alpha \beta}^d$) setting $\epsilon_{\alpha \beta}^d=0$ (setting $\epsilon_{\alpha \beta}^u=0$). The ranges found for $\epsilon_{\alpha \beta}^u$ and  $\epsilon_{\alpha \beta}^d$ turn out to be very similar especially for the elements which are obtained dominantly from atmospheric data for which $N_u\simeq N_d$. For elements that are derived from solar neutrino data ({\it e.g.,} $\epsilon_{ee}^d-\epsilon_{\mu\mu}^d$ and $\epsilon_{ee}^u-\epsilon_{\mu\mu}^u$)  the corresponding ranges are slightly different as the Sun is mostly composed of proton so $N_u/N_d\simeq 2$. We take $\epsilon_{\alpha \beta}\simeq 3\epsilon_{\alpha \beta}^u\simeq 3\epsilon_{\alpha \beta}^d $ to translate the bounds reported in Ref \cite{Maltoni} on $\epsilon_{\alpha \beta}^u$ and $\epsilon_{\alpha \beta}^d $ into bounds on $\epsilon_{\alpha \beta}$ which is the combination relevant for neutrino propagation in earth. Two solutions have been found in Ref \cite{Maltoni}. One of them is consistent with standard interactions and constrains $\epsilon$ to the following range at 1$\sigma$ C.L:
 \begin{eqnarray} \label{boundsonE}
 \begin{matrix}
 |\epsilon_{e\mu}|&<& 0.16\\
 |\epsilon_{e\tau}|&<& 0.26\\
 |\epsilon_{\mu\tau}|&<& 0.02
\end{matrix}
 \end{eqnarray}
 and
  \begin{eqnarray} \label{boundsonE1}
 \begin{matrix}
 -0.018&<&\epsilon_{\tau\tau}-\epsilon_{\mu \mu}&<& 0.054\\
 0.35&<&\epsilon_{ee}-\epsilon_{\mu \mu}&<& 0.93
 %-0.54<\epsilon_{e\mu}<0.24 ~~~~~~~& ~~~~~~~ (-0.96<\epsilon_{e\mu}<0.66) \\
  %-0.84<\epsilon_{e\tau}<0.84 ~~~~~~~ &  ~~~~~~~(-2.4<\epsilon_{e\tau}<1.8) \\
  %-0.06<\epsilon_{\mu\tau}<0.06  ~~~~~~~& ~~~~~~~ (-0.18<\epsilon_{\mu\tau}<0.18) \\
 %0<\epsilon_{ee}-\epsilon_{\mu\mu}<3.06 ~~&~~~~ (-0.54 <\epsilon_{ee}-\epsilon_{\mu\mu}<4.26) \\
 %-0.06<\epsilon_{\tau\tau}-\epsilon_{\mu\mu}<0.18 ~~~~~~&~~~~ (-0.18 <\epsilon_{\tau\tau}-\epsilon_{\mu\mu}<1.2)
 \end{matrix}
 \end{eqnarray}
 The other solution is the famous LMA-Dark solution with $\theta_{12}> 45^\circ$ and  $\epsilon_{\mu \mu}-\epsilon_{ee}\sim 1$ \cite{Mariam}. As shown in \cite{us}, this solution can be tested by intermediate reactor experiments JUNO and RENO-50.

  %at $1\sigma$ ($3\sigma$) C.L.:
  %\begin{eqnarray} \label{boundsonEDARK}
 %\begin{matrix}
 %|\epsilon_{e\mu}|&<& 0.3(0.5)\\
 %|\epsilon_{e\tau}|&<& 0.3(1.2)\\
 %|\epsilon_{\mu\tau}|&<& 0.02(0.09)\\
 %|\epsilon_{\tau\tau}-\epsilon_{\mu \mu}|&<& 0.1(0.6)\\
 %|\epsilon_{ee}-\epsilon_{\mu \mu}|&=& 3(4)
% -0.54<\epsilon_{e\mu}<0.6 ~~~~~~~& ~~~~~~~ (-0.96<\epsilon_{e\mu}<1.02) \\
 % -0.90<\epsilon_{e\tau}<0.84 ~~~~~~~ &  ~~~~~~~(-2.4<\epsilon_{e\tau}<2.4) \\
  %-0.06<\epsilon_{\mu\tau}<0.06  ~~~~~~~& ~~~~~~~ (-0.18<\epsilon_{\mu\tau}<0.18) \\
 %-7.14<\epsilon_{ee}-\epsilon_{\mu\mu}<-4.86 ~~&~~~~ (-8.4 <\epsilon_{ee}-\epsilon_{\mu\mu}<-4.08) \\
 %-0.18<\epsilon_{\tau\tau}-\epsilon_{\mu\mu}<0.18 ~~~~~~&~~~~ (-1.14 <\epsilon_{\tau\tau}-\epsilon_{\mu\mu}<1.2)
 %\end{matrix}
 %\end{eqnarray}
 We can always add a matrix proportional to the unit matrix $I_{3\times 3}$ to the Hamiltonian in Eq.~(\ref{Hmat}) without changing the neutrino oscillation pattern.
 That is why from neutrino oscillation data only a bound on the difference of diagonal elements of $\epsilon$ ({\it i.e.,} $\epsilon_{\mu \mu}-\epsilon_{\tau \tau}$ and/or
  $\epsilon_{\mu \mu}-\epsilon_{ee}$) can be derived. For consistency we set $\epsilon_{\mu \mu}=0$ throughout our analysis. Hermiticity of $H_{mat}$ implies that the diagonal elements of $\epsilon$ are real but they can be positive or negative.
  The off-diagonal elements of $\epsilon$ can be in general complex. There is no observational constraint on the phases of $\epsilon_{e\mu}$, $\epsilon_{e\tau}$ and $\epsilon_{\mu \tau}$.
    %To derive the bounds in Eqs. (\ref{boundsonE},\ref{boundsonEDARK}), Ref \cite{Maltoni} take them real. To perform our analysis, we vary $|\epsilon_{\alpha \beta}|$
    %in the allowed range compatible with $3\sigma$ range in  Eqs. (\ref{boundsonE},\ref{boundsonEDARK}) and allow the phases of $\epsilon_{\alpha \beta}$ with $\alpha \ne \beta$
    %to vary in the range $(0,2\pi)$.

    As seen from Eqs. (\ref{boundsonE},\ref{boundsonE1}), there are already strong bounds on $|\epsilon_{\mu\tau}|$ and on $|\epsilon_{\mu\mu}-\epsilon_{\tau\tau}|$.
    We can write $|\epsilon_{\mu\mu}-\epsilon_{\tau\tau}|,|\epsilon_{\mu \tau}|\stackrel{<}{\sim} \sin \theta_{13}$. On the other hand, up to $O(s_{13}^2\epsilon, s_{13}\epsilon^2
    ,\epsilon^3)$, $P(\nu_\mu \to \nu_e)$ does not depend on  $|\epsilon_{\mu\mu}-\epsilon_{\tau\tau}|$ and $|\epsilon_{\mu \tau}|$ \cite{Kopp:2007ne,Danny}.
    Our numerical analysis show that
    $\epsilon_{\mu \tau}$ or $|\epsilon_{\mu \mu}-\epsilon_{\tau \tau}|$ do not cause any degeneracy in the determination of $\delta$.
This is expected as the appearance mode dominates the $\delta$ determination.
    Numerical calculations also confirm this claim. However, nonzero $\epsilon_{e\mu}$, $\epsilon_{e\tau}$ and $\epsilon_{ee}-\epsilon_{\mu\mu}$ can interfere with the determination
    of $\delta$ \cite{Danny,Poonam}. We study how MOMENT can help to solve the degeneracies caused by turning on nonzero $\epsilon_{e\mu}$, $\epsilon_{e\tau}$ and
     $\epsilon_{ee}$. We calculate the oscillation probabilities numerically. As expected, the oscillation pattern for nonzero $\epsilon_{ee}=\epsilon$ and
     $\epsilon_{\mu\mu}=\epsilon_{\tau\tau}=0$ is the same as for nonzero $\epsilon_{\mu\mu}=\epsilon_{\tau\tau}=\epsilon$ and
     $   \epsilon_{ee}=0$. To perform our analysis, we set true values of $|\epsilon|$ to zero and treat uncertainties in $|\epsilon|$ with pull method \cite{Fogli,Huber:2004ka,Huber:2007ji}.
     We marginalize over phases of $\epsilon_{e\mu}$, $\epsilon_{e\tau}$ and $\epsilon_{\mu\tau}$.
 %%%%%%%%%%%%%%%%%%%%%%%%%%%%%%%%%%%%%%%%%%%%%%%%%%%%%%%%%
\section{ Characteristics of MOMENT, T2K and NO$\nu$A \label{char}}
%%%%%%%%%%%%%%%%%%%%%%%%%%%%%%%%%%%%%%%%%%%%%%%%%%
The proposal of the MOMENT experiment is still in a early stage and its details have not been completely fixed. To  make a comparison we will assume characteristics for the
MOMENT setup similar to those in \cite{Pilar-Moment}. We take $L=150$ km and a Gd-doped water Cherenkov detector with  fiducial mass of 500 kton.
The source can run in two modes: 1) muon  mode, $\mu^- \to e^- \bar{\nu}_e\nu_\mu$; 2) antimuon  mode,   $\mu^+ \to e^+ {\nu}_e\bar{\nu}_\mu$.
The power and spectrum of two modes are taken to be the same.
The energy spectrum of neutrinos at source is taken from \cite{wang}. The peak energy lies in around 150 MeV and the maximum energy is around 700 MeV. At this energy range, the dominant interaction modes
are quasi-elastic interactions:
$$\nu_e+n\to p+e^-  \ \ \ \ \ \ \ \bar{\nu}_\mu+p \to n+\mu^+$$ and
$$\bar{\nu}_e+p\to n+e^+  \ \ \ \ \ \ \ \nu_\mu+n \to p+\mu^-.$$
The final neutron can be captured on Gd which provides a method to distinguish neutrinos from antineutrinos. We shall assume that Charge Identification (CI) is 80 \%
 which is although  relatively optimistic but is not  unrealistic \cite{CI}. The charge misidentification is the main source of background. Another important source of background is atmospheric neutrinos.
 By sending the beam in bunches, the atmospheric neutrino background can be reduced by a factor called Suppression Factor (SF).
% Similarly to \cite{Pilar-Moment},
In most of our analysis, we take SF=0.1 \%. We will then study the dependence of results on SF.
Another non-negligible source of background is neutral current interactions \cite{Pilar-Moment} which we take into account.  Since the energies at MOMENT are low, pion production
 will not be a problem. Moreover, since the water Cherenkov detectors enjoy very good flavor identification, background from the flavor misidentification will be negligible.  We take the backgrounds similar to those in
 \cite{Pilar-Moment}. We take the spectrum of neutrinos at the source from \cite{wang}. We  take the unoscillated neutrino flux of each flavor  mode  at the detector equal to  $4.7\times 10^{11}$ m$^{-2}$~year$^{-1}$. We assume five years of data taking in each muon and anti-muon modes. Uncertainties of (unoscillated) flux normalization of $\nu_e$ and $\bar{\nu}_\mu$ are taken to be the same and equal to 5 \%. Similarly we take an uncertainty of 5 \% in flux renormalization of
  $\bar{\nu}_e$ and ${\nu}_\mu$ in the muon decay mode but the uncertainties of fluxes of muon and anti-muon decay modes are taken to be uncorrelated. For neutrino energy resolution, we include migration matrix similar to Ref. \cite{BurguetCastell:2005pa}.
  For cross section of quasi-elastic Charged Current (CC) interactions, we use the results of Ref. \cite{Paschos:2001np,Messier:1999kj}. The efficiencies of various signal modes are taken from \cite{MEMphys}.

 For studying the synergies between experiments, we also forecast the final results of T2K and NO$\nu$A. We assume 2 (6) years of data taking in neutrino (antineutrino)
  mode for T2K and 3 years of data taking in each neutrino and antineutrino mode for NO$\nu$A. In our analysis of  T2K and NO$\nu$A,  we take into account all the
   electron and muon appearance and disappearance channels.
   The flux of T2K is taken from Ref.~\cite{Itow:2001ee}. The energy resolution for T2K is set equal to $85~{\rm MeV}$ uniformly for all energies.
   The energy range is between $0.4$ to $1.2$~GeV. The baseline is 295 km.
   $5\%$ and $2.5\%$ normalization uncertainty are considered for appearance mode and disappearance signal mode, respectively. Free normalization is considered for quasi elastic events. Background sources include  lepton flavor misidentification, neutral current events, charge misidentification and intrinsic background. For the  backgrounds of the disappearance channels, we take a $20\%$ normalization uncertainty   and for  backgrounds of appearance channels, we take  an uncertainty of $5\%$.  The calibration error is considered equal to $0.01\%$ for both signal and background.
          Simulating the T2K experiment, we take its features as described in  Ref. \cite{Itow:2001ee,Huber:2002mx} and its systematics as Ref. \cite{Ishitsuka:2005qi}.

    The energy range of NO$\nu$A experiment is from 1 to 3.5 GeV and the baseline is 812 km. The energy resolution is equal to 10$\%\sqrt{E}$ and 5$\%\sqrt{E}$ for electron neutrino and muon neutrino, respectively. A  normalization uncertainty of 5$\%$ is considered for signal and background. The calibration error is 2.5$\%$. Backgrounds include neutral current interaction, lepton flavor misidentification and the intrinsic background.
      Simulating the NO$\nu$A experiment, we take the  features  of appearance and disappearance channels  as described in Ref. \cite{Ayres:2004js} and  in Ref. \cite{Yang_2004}, respectively.

    {The simulated number of events  for the appearance and disappearance channels of MOMENT, T2K and NO$\nu$A experiment are shown in Table \ref{table}. We take the oscillation parameters from nu-fit \cite{Gonzalez-Garcia:2014bfa} and set $\delta=270^\circ$. Notice the number of events includes both signal and background.}

 We perform our analysis using GLoBES \cite{Huber:2004ka,Huber:2007ji}. The neutrino oscillation probabilities are calculated  using the numerical diagonalization method discussed in \cite{diagonal} (see also \cite{Kopp:2007ne,Kopp:2006wp}).
For matter density profile, we use PREM with 5\% uncertainties  \cite{PREM}.
The neutrino mass and mixing parameters are taken from \cite{Gonzalez-Garcia:2014bfa}. To treat all the uncertainties we use the pull method.

\begin{table}
\begin{center}
    \begin{tabular}{ | l | l | l | l | l | p{5cm} |}
    \hline
    Experiment & neutrino~ mode &  neutrino~ mode &  antineutrino~ mode &  antineutrino~ mode \\
    ~ & $\nu_\mu$ & $\nu_e$ & $\bar{\nu}_\mu$ & $\bar{\nu}_e$ \\ \hline
    T2K & 248 & 58 & 255 & 31 \\ \hline
    NO$\nu$A & 1326 & 142 & 502 & 37 \\ \hline \hline
    MOMENT        & $\nu_\mu$ & $\bar{\nu}_\mu$  & $\nu_e$ & $\bar{\nu}_e$   \\
    \hline
    $\bar{\nu}_\mu$, $\nu_e$ beam & 941 & 2054 & 21259 & 5544 \\
    \hline
    $\nu_\mu$, $\bar{\nu}_e$ beam & 4954 & 1664 & 3174 & 7549 \\
    \hline
    \end{tabular}
    \caption{\label{table} Number of simulated  events  (signal+background) for T2K  \cite{Itow:2001ee,Huber:2002mx}, NO$\nu$A  \cite{Ayres:2004js,Yang_2004} and MOMENT experiment. The known oscillation parameters are taken from nu-fit \cite{Gonzalez-Garcia:2014bfa} and the value of $\delta$  is set equal to $270^\circ$.}
\end{center}
\end{table}

%%%%%%%%%%%%%%%%%%%%%%%%%%%%%%%%%%%%%%%%%%
\section{Results\label{resul}}
%%%%%%%%%%%%%%%%%%%%%%%%%%%%%%%%%%%%%%%%%%%%%%
In this section we discuss our results which are shown in Figs 1-6. Drawing all these figures, we set the true values of neutrino mass and mixing parameters equal to
their best fit values  \cite{Gonzalez-Garcia:2014bfa}. The uncertainties of those parameters that are not shown on the axes are taken from  \cite{Gonzalez-Garcia:2014bfa}
and treated by pull-method. The true values of $\epsilon$ are set to zero. As explained in sec \ref{theo}, the dependence of neutrino oscillation patterns  on diagonal elements of $\epsilon$ is only through differences $\epsilon_{ee}-\epsilon_{\mu \mu}$ and $\epsilon_{\tau \tau}-\epsilon_{\mu \mu}$. We therefore fix $\epsilon_{\mu \mu}=0$.

\begin{figure}
\begin{center}
\subfigure[]{\includegraphics[width=0.49\textwidth]{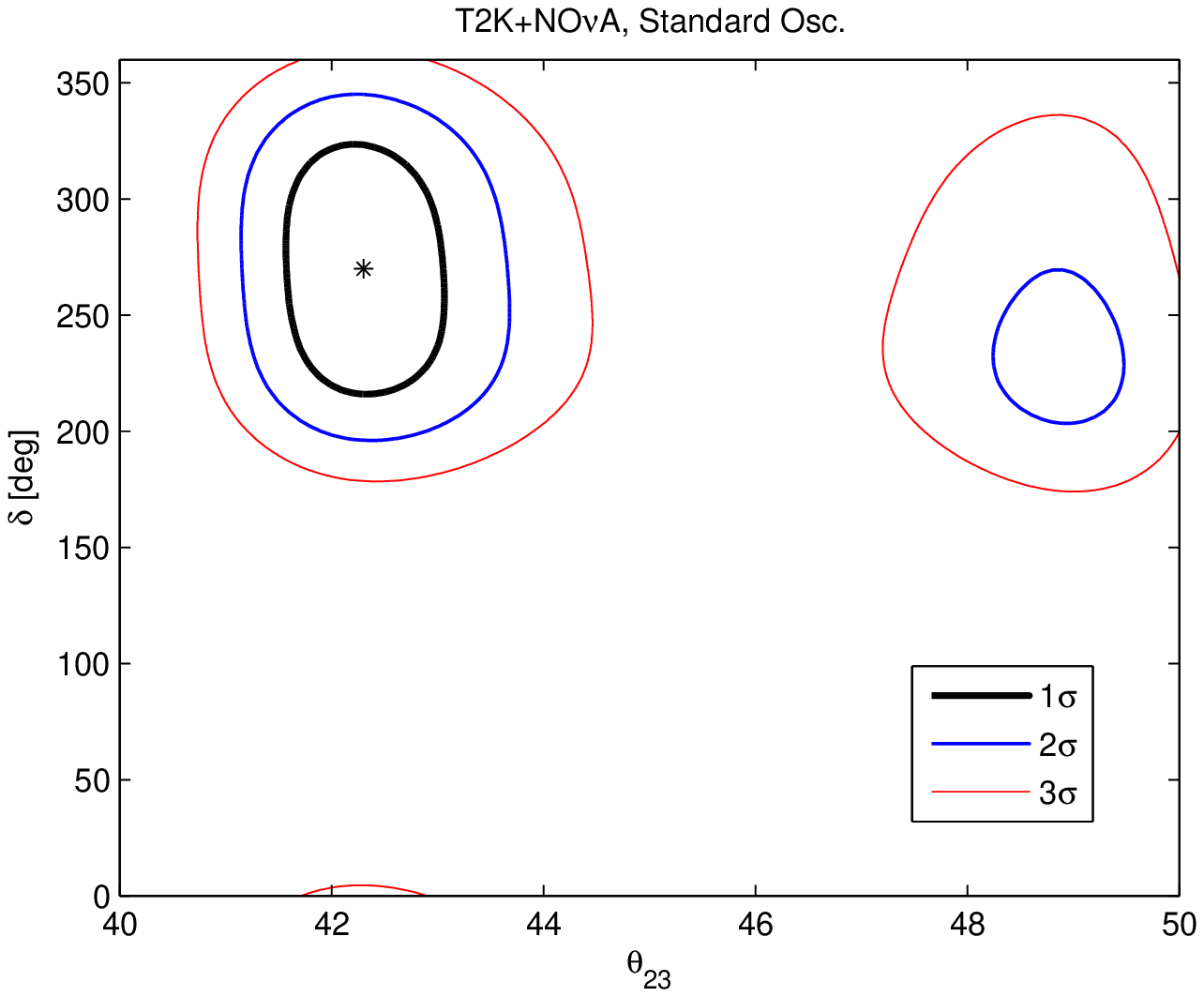}}
\subfigure[]{\includegraphics[width=0.49\textwidth]{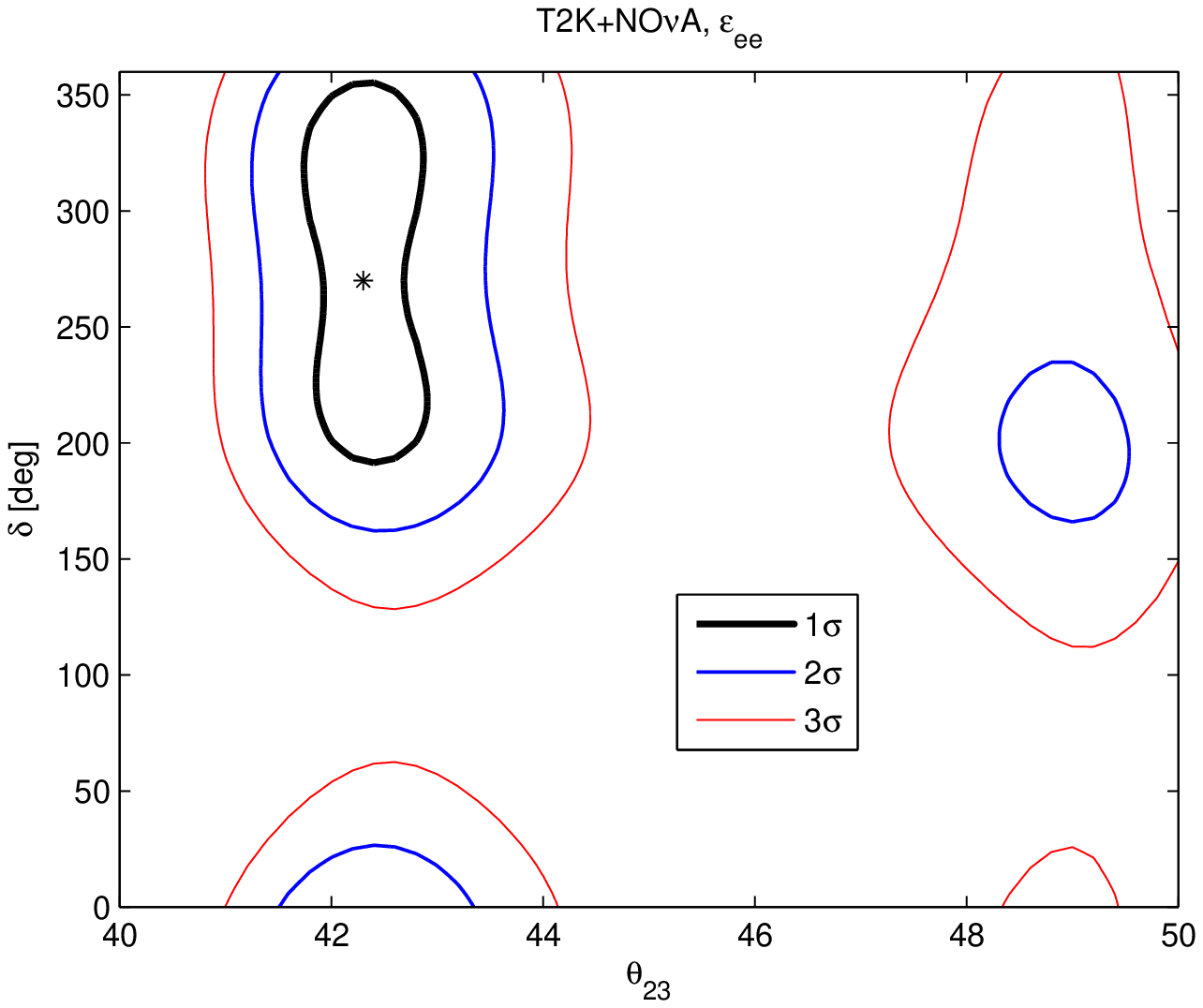}}
\subfigure[]{\includegraphics[width=0.49\textwidth]{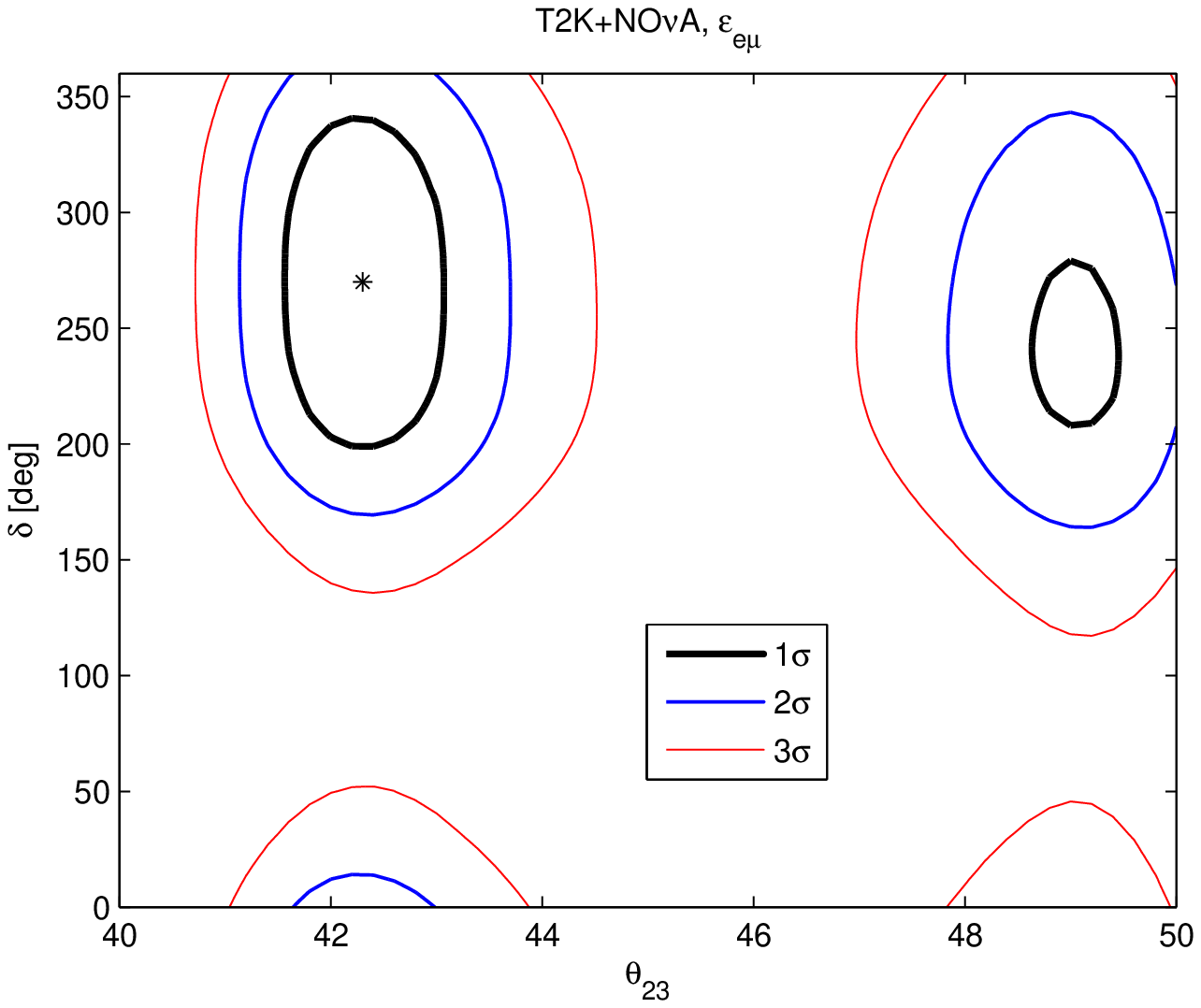}}
\subfigure[]{\includegraphics[width=0.49\textwidth]{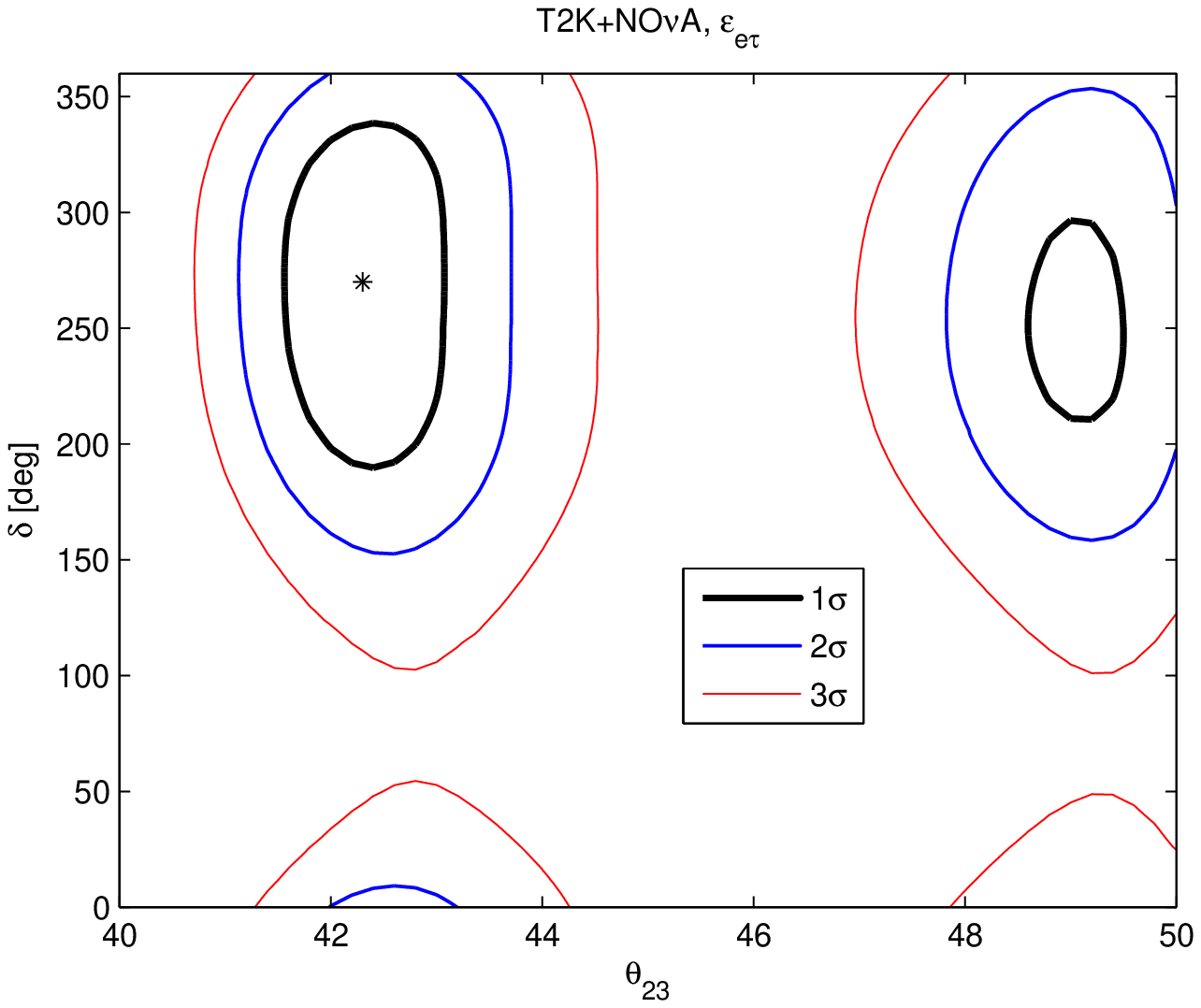}}
\end{center}
\vspace{2cm} \caption[]{Projected  combined sensitivity of NO$\nu$A and T2K on $\delta-\theta_{23}$. The true values of neutrino mass parameters are marked by a star and are set  to their present best fit values \cite{Gonzalez-Garcia:2014bfa}. Both appearance and disappearance modes are taken into account. Fig (a) shows the projected sensitivity assuming no NSI.  In Fig (b), the present 1$\sigma$ uncertainty of $\epsilon_{ee}$ \cite{Maltoni} is taken into account. In Figs (c) and (d), the present 1$\sigma$ uncertainties of respectively  $\epsilon_{e\mu}$ and $\epsilon_{e\tau}$ \cite{Maltoni} are taken into account, varying their phases in $(0,2\pi)$. }
\label{T2KNOvA}
\end{figure}
Fig. \ref{T2KNOvA} shows the effects of turning on NSI on determination of $\delta-\theta_{23}$ by the current long baseline experiments NO$\nu$A and T2K.
We assume the normal mass ordering. Moreover we assume that the ordering is known.   The true values are shown by a star: $\delta=270^\circ$ and $\theta_{23}=42.3^\circ$. In Fig \ref{T2KNOvA}-a, all the NSI are turned off. This figure confirms the results shown in Fig \ref{T2KNOvA}-a of \cite{Pilar-Moment}.
In Figs \ref{T2KNOvA}-b, \ref{T2KNOvA}-c and \ref{T2KNOvA}-d, the parameters $\epsilon_{ee}$, $\epsilon_{e\mu}$ and $\epsilon_{e\tau}$ are respectively allowed to vary within the present 1 $\sigma$ C.L.
intervals shown in  Eqs. (\ref{boundsonE},\ref{boundsonE1}). The phases of $\epsilon_{e\mu}$ and $\epsilon_{e\tau}$ are allowed to vary in $(0, 2\pi)$.
We observe that turning on $\epsilon_{e\mu}$ or $\epsilon_{e\tau}$, T2K and NO$\nu$A lose their power to determine the octant of $\theta_{23}$ even at $1\sigma$ C.L.
 \begin{figure}
\begin{center}
\subfigure[]{\includegraphics[width=0.49\textwidth]{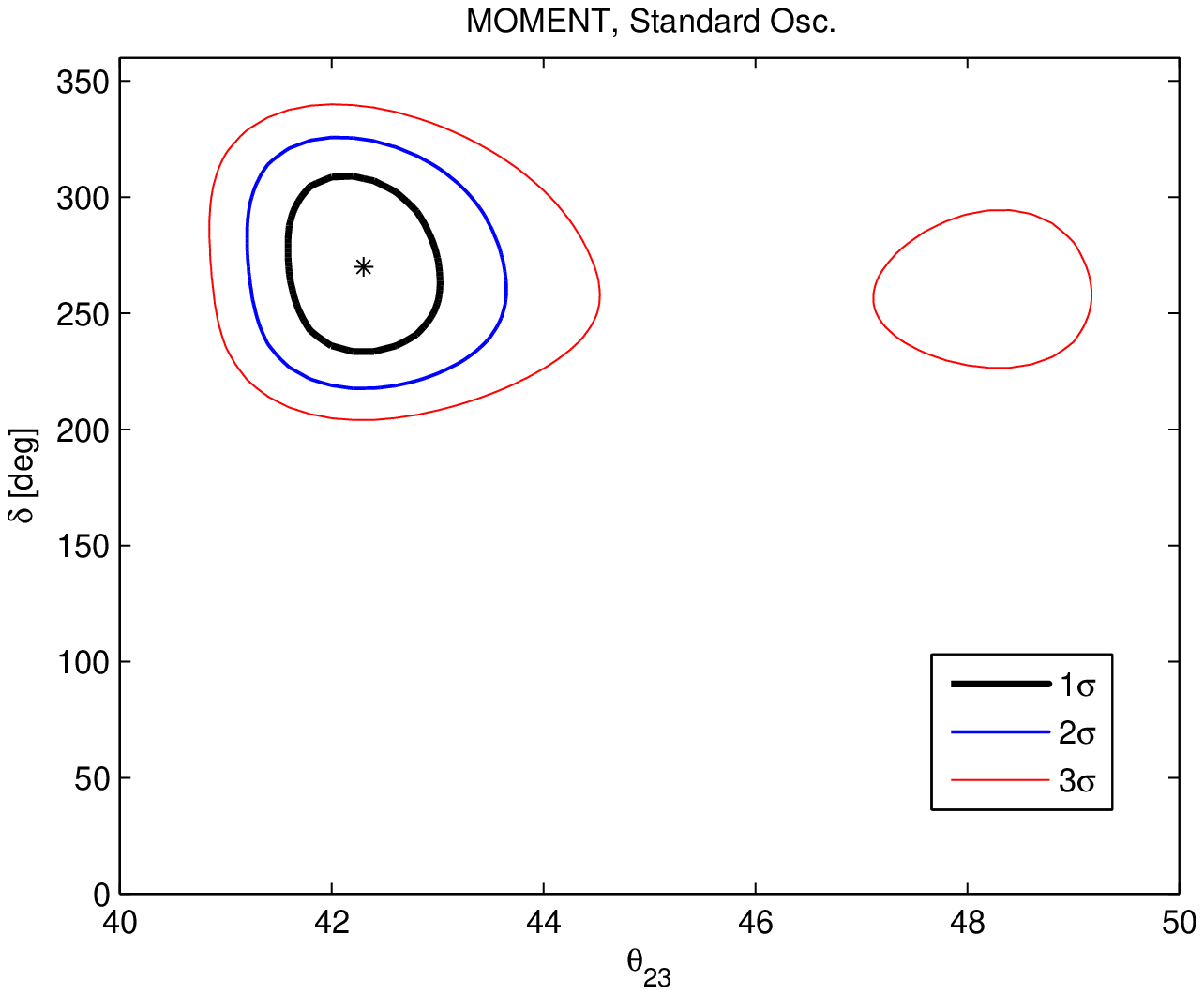}}
\subfigure[]{\includegraphics[width=0.49\textwidth]{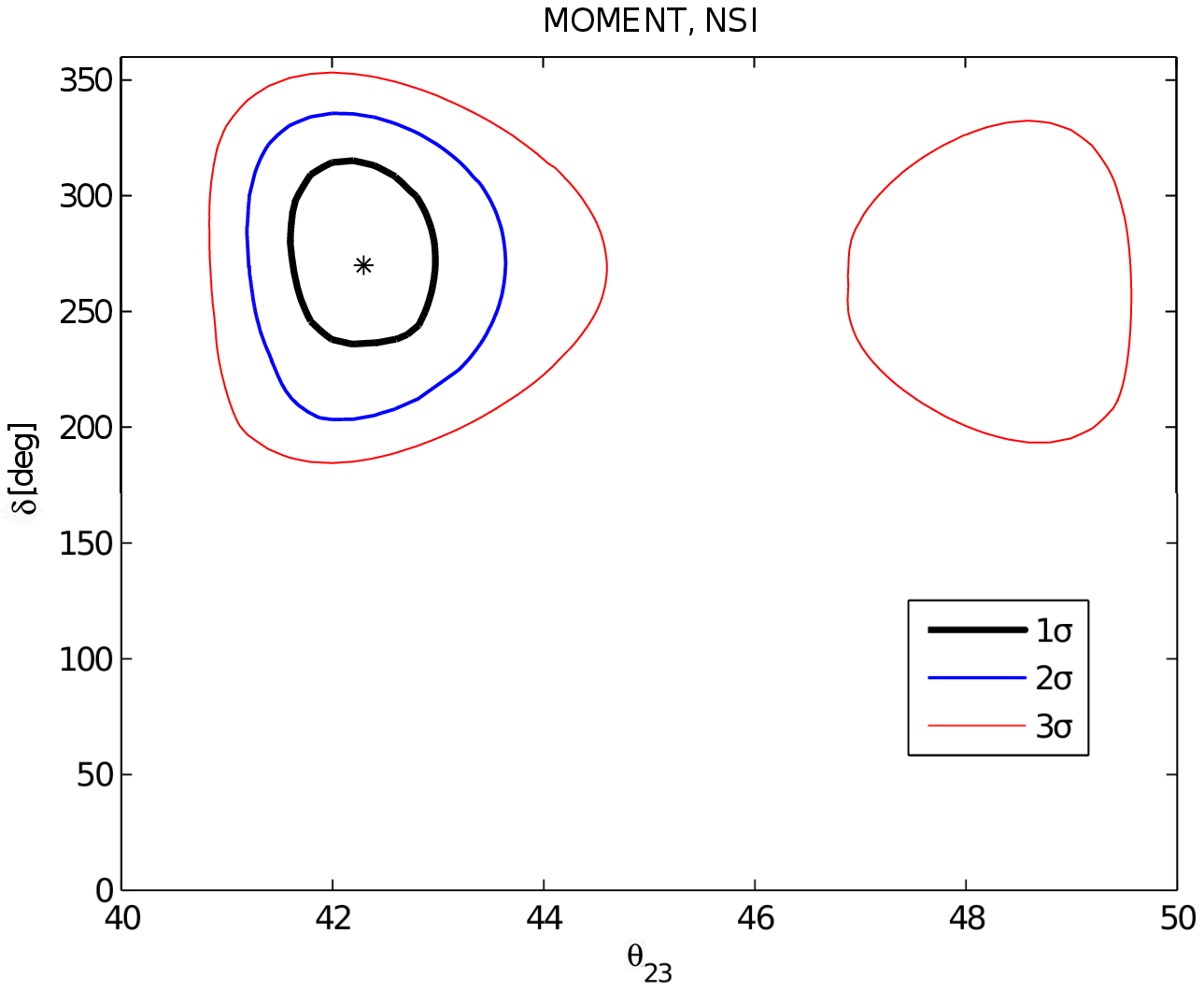}}
\subfigure[]{\includegraphics[width=0.49\textwidth]{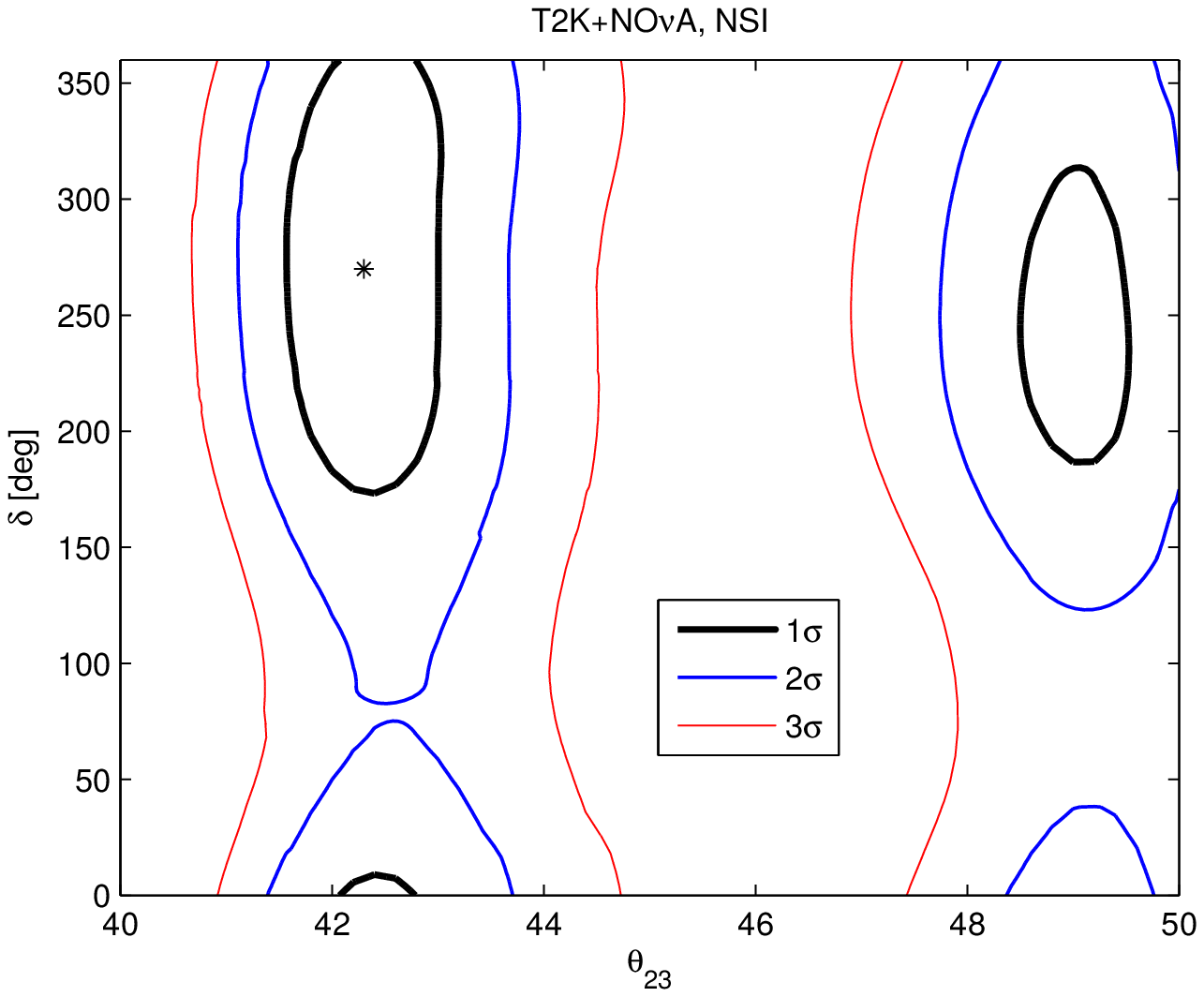}}
\subfigure[]{\includegraphics[width=0.49\textwidth]{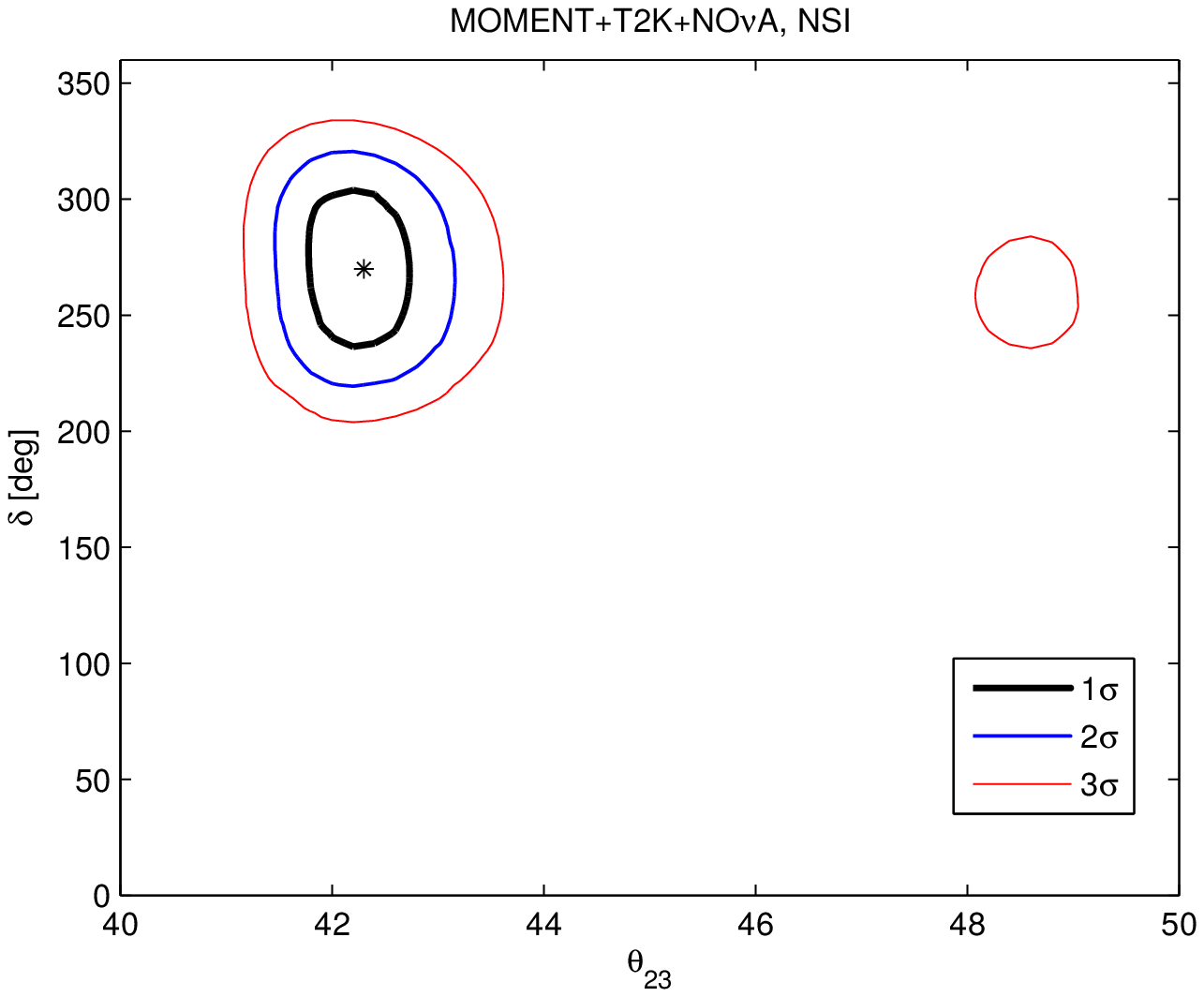}}
\end{center}
\vspace{2cm} \caption[]{Projected sensitivity of MOMENT, NO$\nu$A and T2K on $\delta-\theta_{23}$. The true values of neutrino mass parameters  are marked by a star and   are set to their present best fit values \cite{Gonzalez-Garcia:2014bfa}.  Both appearance and disappearance modes are taken into account.  SF for MOMENT is taken equal to 0.1 \%. Fig (a) shows the sensitivity of MOMENT without NSI. In Figs  (b), (c) and (d), all values of $\epsilon$ are allowed to vary within their present 1$\sigma$ uncertainty limits \cite{Maltoni}. Fig (b) shows the sensitivity of the MOMENT experiment alone. Fig (c) shows the combined sensitivity of the NO$\nu$A and T2K experiments and Fig (d) shows the combined sensitivity of all three experiments. }
\label{MOMENT}
\end{figure}

Fig. \ref{MOMENT} demonstrates how MOMENT  can help T2K and NO$\nu$A  to solve the degeneracies induced by turning on NSI. Fig \ref{MOMENT}-a shows constraints that the MOMENT experiment can put on $\delta$
and $\theta_{23}$ when there is no NSI.  This figure is in agreement with the results of \cite{Pilar-Moment}. In Fig. \ref{MOMENT}-b, we allow all elements of  $\epsilon$
to vary within the range shown in  Eqs. (\ref{boundsonE},\ref{boundsonE1}) and the phases of off-diagonal elements of  $\epsilon$ are taken in the range $[0,2\pi]$. As expected the uncertainties only slightly
increase compared to Fig \ref{MOMENT}-a because the MOMENT experiment is not very sensitive to the matter effects (neither standard nor non-standard). Fig \ref{MOMENT}-c shows $\delta-\theta_{23}$
contours by NO$\nu$A and T2K allowing the $\epsilon$ elements and their phases vary within the aforementioned range. As seen from this figure
at 3 $\sigma$ C.L. all values of $\delta$ are allowed.  This confirms the result of \cite{Huber} that the effects
of $\delta =270^\circ$ can be mimicked with NSI even when CP is conserved ({\it i.e.,} $\delta=0$ or $180^\circ$ and Im($\epsilon_{\alpha \beta}$)=0). Fig. \ref{MOMENT}-d demonstrates the improvement once we add the data from MOMENT. As seen from this figure, with the help of
MOMENT, CP-violation can be established for $\delta=270^\circ$ even when we allow all the elements of $\epsilon$ to vary. Remember that this is a task that cannot be achieved
even by DUNE \cite{Pilar-DUNE,Danny}.
 \begin{figure}
\begin{center}
\subfigure[]{\includegraphics[width=0.49\textwidth]{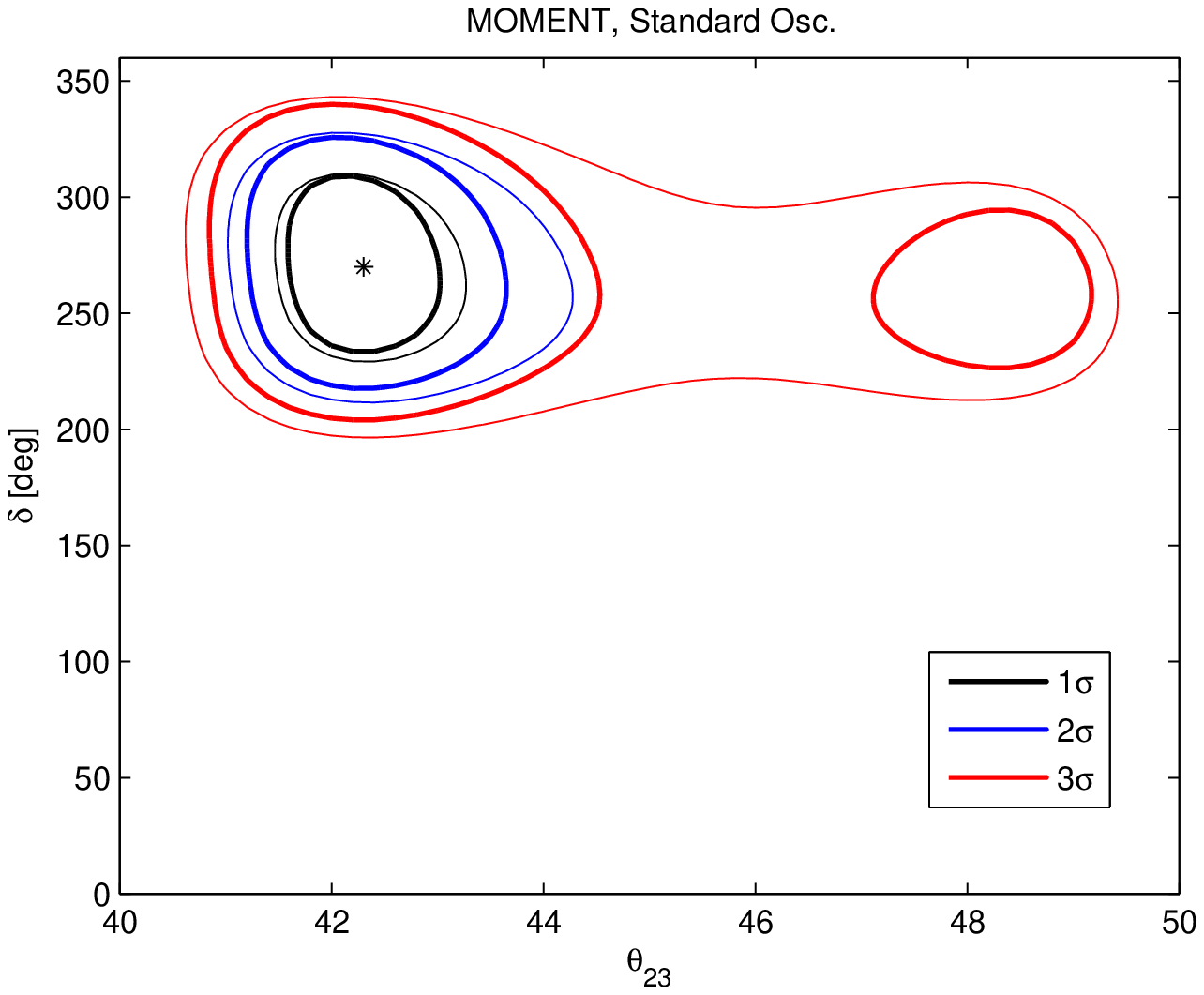}}
\subfigure[]{\includegraphics[width=0.49\textwidth]{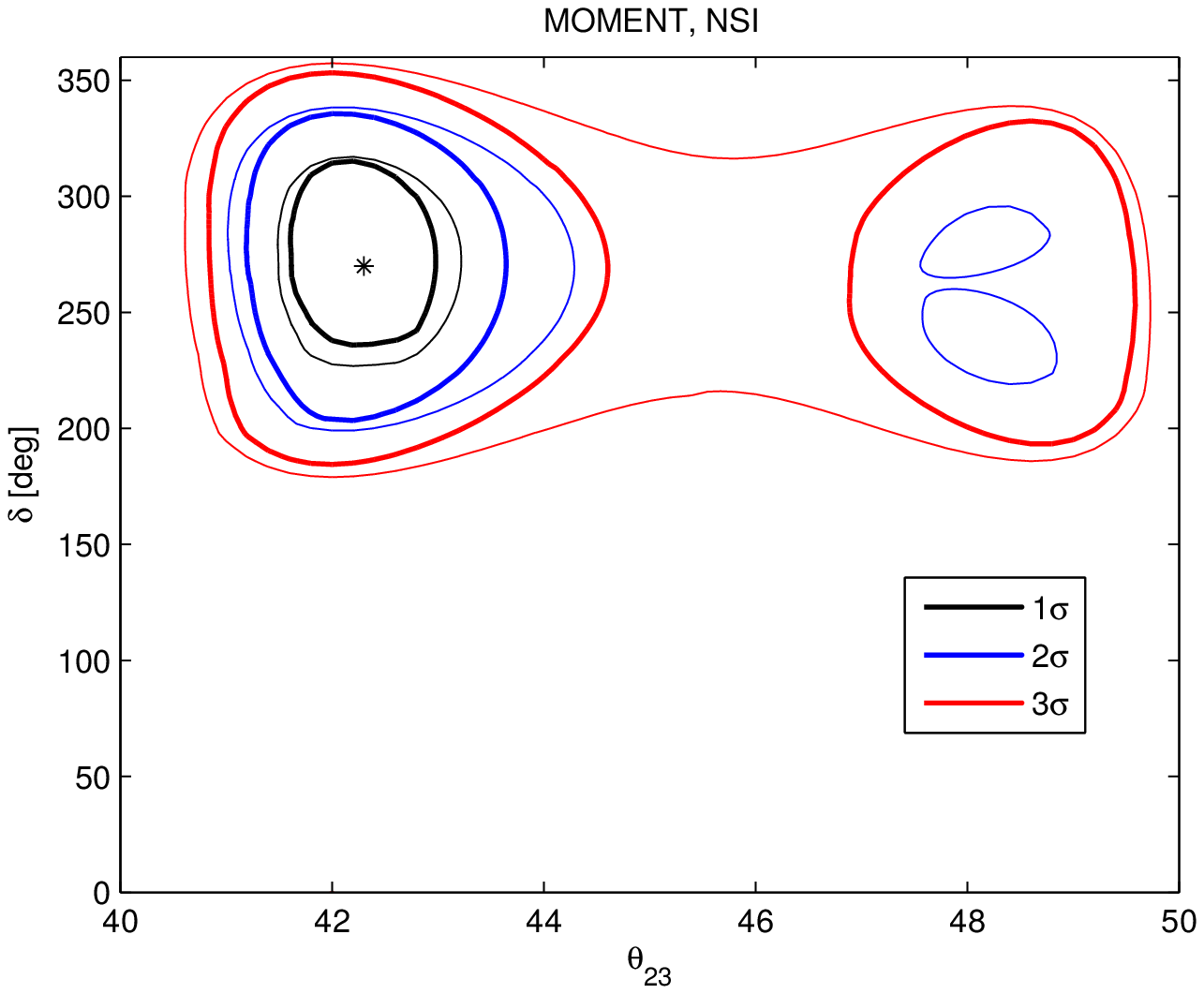}}
\end{center}
\vspace{2cm} \caption[]{Projected sensitivity of the MOMENT experiment on $\delta-\theta_{23}$ for different background  suppression factors SF=0.1 \% and 10\%, respectively shown with thick and thin lines. The true values of neutrino mass parameters are marked by a star and are set to their present best fit values \cite{Gonzalez-Garcia:2014bfa}.  Fig (a) shows the projected sensitivity assuming no NSI.  In Fig (b), the present 1$\sigma$ uncertainty of all values of $\epsilon$ \cite{Maltoni} are taken into account.}
\label{SF}
\end{figure}

Fig \ref{SF} demonstrates the dependence of the sensitivity of the MOMENT experiment on the background Suppression Factor (SF). As expected for larger background ({\it i.e.,} increasing SF), the uncertainty on $\theta_{23}$ and $\delta$ increases. From these figures, we observe that  with SF=10 \%, MOMENT  will not be able to  tell whether $\theta_{23}$ is maximal or not at 3 $\sigma$. However determination of $\delta$ is not so sensitive to SF for SF better than 10\%. That is for the purpose of determining $\delta$, background suppression factor below 10 \% is not necessary. This result is in agreement with the results of \cite{Pilar-Moment} shown in its Fig 2 for standard oscillation. Comparing Fig \ref{SF}-a and \ref{SF}-b, we observe that these results are robust against turning on NSI. For SF worse than 10 \%, the background will be problematic for the $\delta$ determination \cite{Pilar-Moment}.

\begin{figure}
\begin{center}
\subfigure[]{\includegraphics[width=0.49\textwidth]{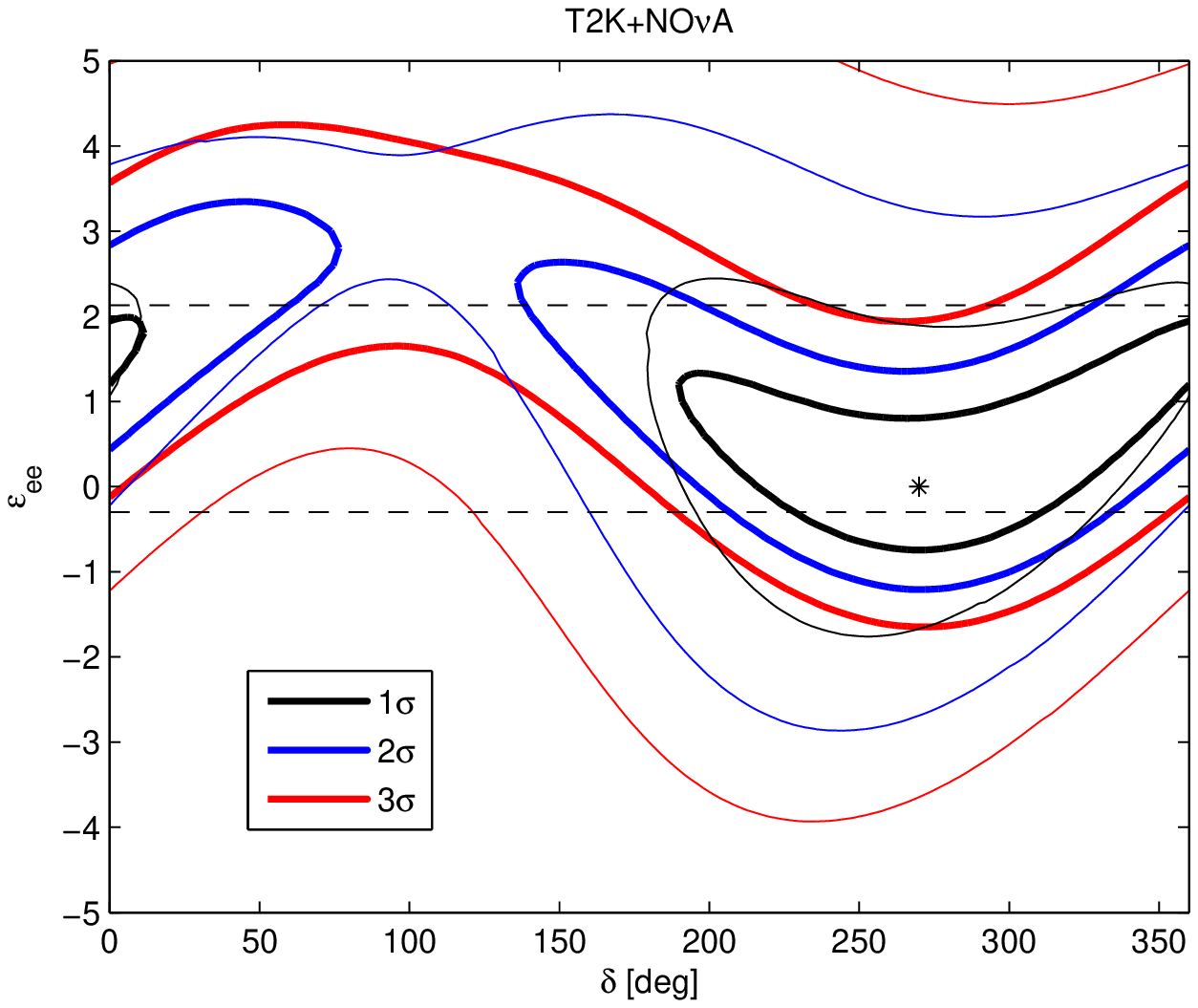}}
\subfigure[]{\includegraphics[width=0.49\textwidth]{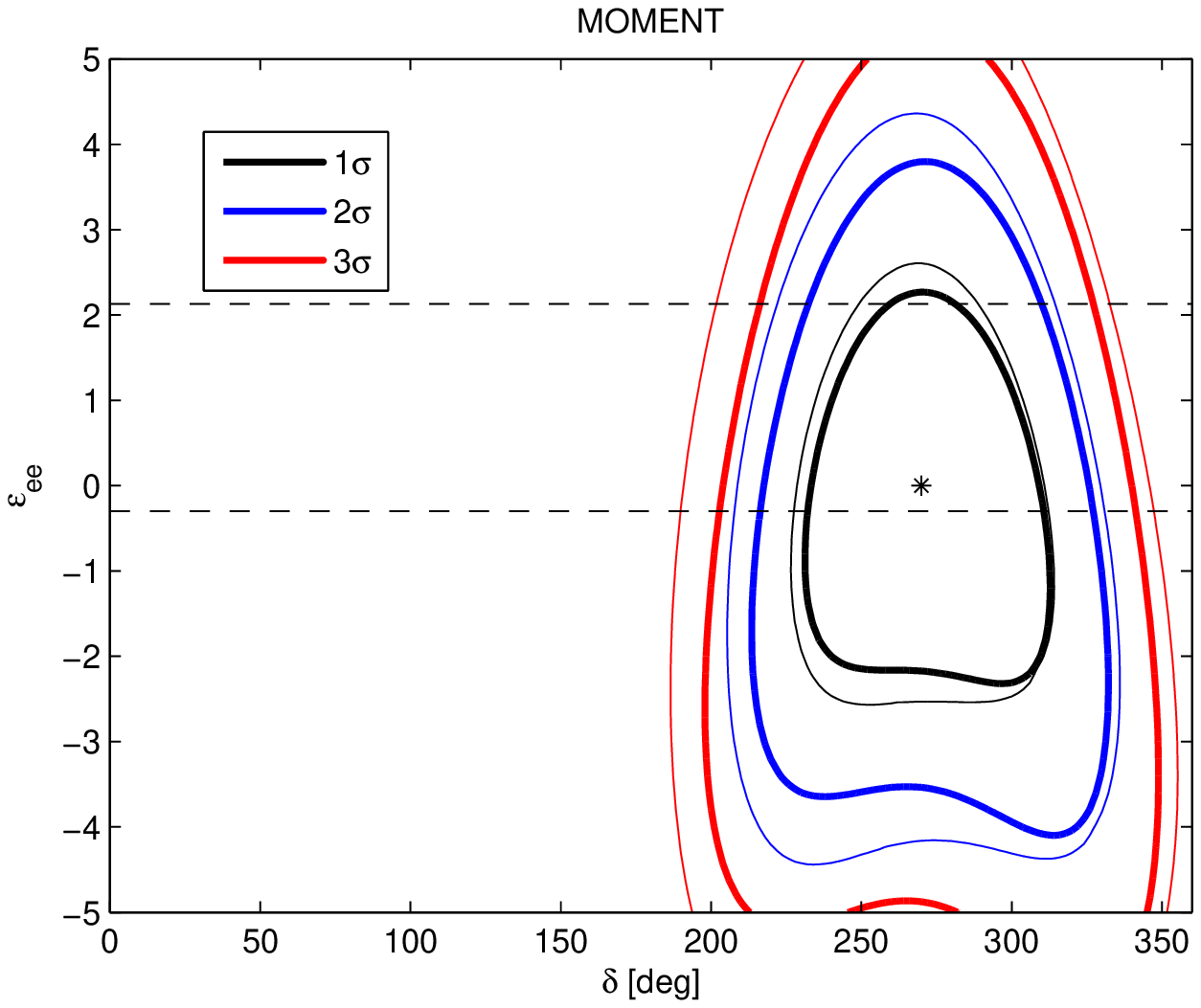}}
\subfigure[]{\includegraphics[width=0.49\textwidth]{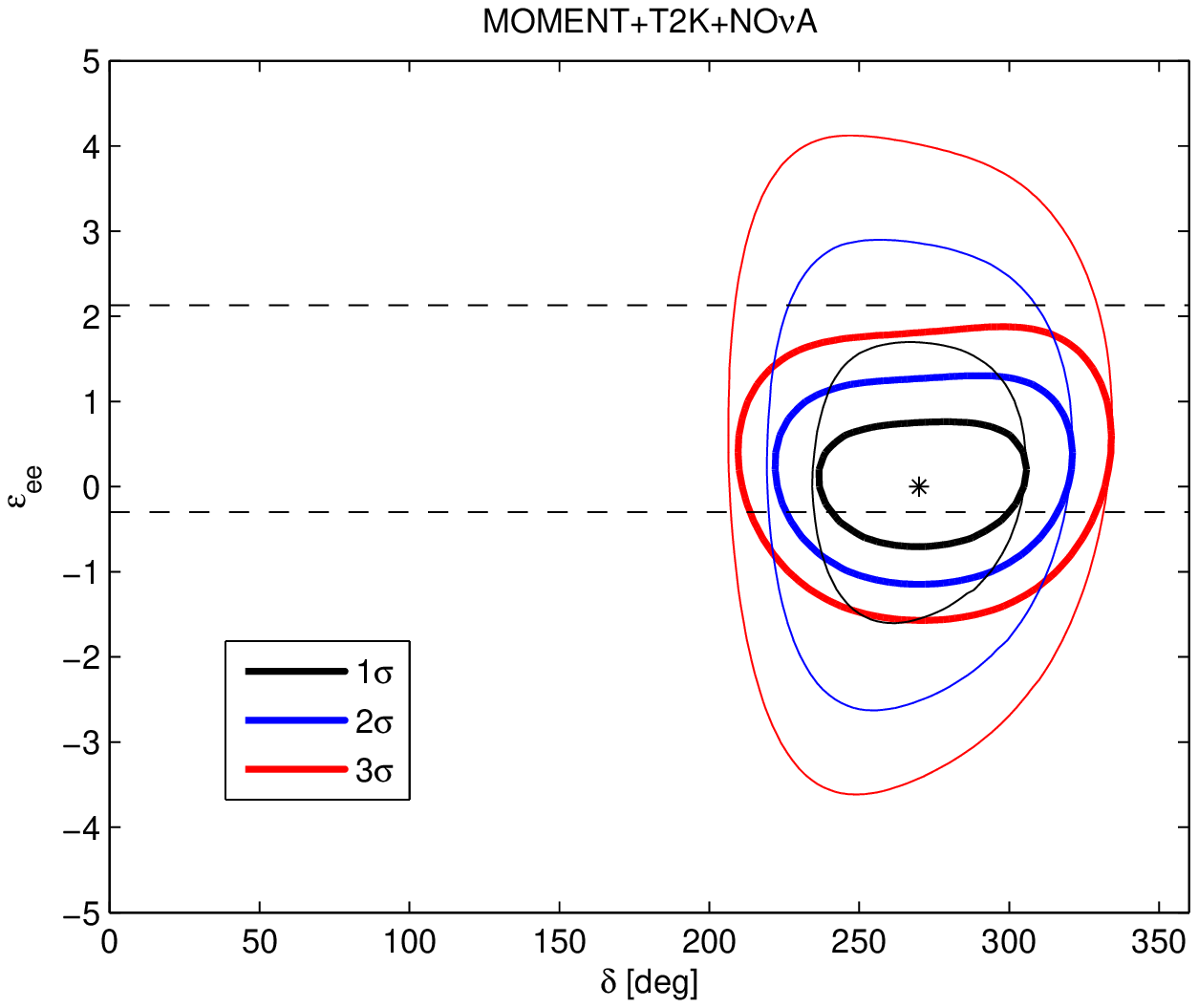}}
\end{center}
\vspace{2cm} \caption[]{Projected sensitivity of MOMENT, NO$\nu$A and T2K on $\epsilon_{ee}-\delta$. The true values of neutrino mass parameters (marked by a star)  are set to their present best fit values \cite{Gonzalez-Garcia:2014bfa}.   Both appearance and disappearance modes are taken into account. The horizontal dashed lines show the present 3$\sigma$ range of $\epsilon_{ee}$.   SF for MOMENT is taken equal to 0.1 \%. { In drawing thick lines, all the $\epsilon_{\alpha \beta}$ except $\epsilon_{ee}$ are fixed to zero but when drawing the thin lines, we have allowed $|\epsilon_{e\tau}|$ and its phase to vary within the uncertainties.} Fig (a) shows the combined sensitivity of the NO$\nu$A and T2K experiments. Fig  (b) shows the sensitivity of the MOMENT experiment alone.  Fig (c) shows the sensitivity of all three experiments combined.}
\label{ee}
\end{figure}
 \begin{figure}
\begin{center}
\subfigure[]{\includegraphics[width=0.49\textwidth]{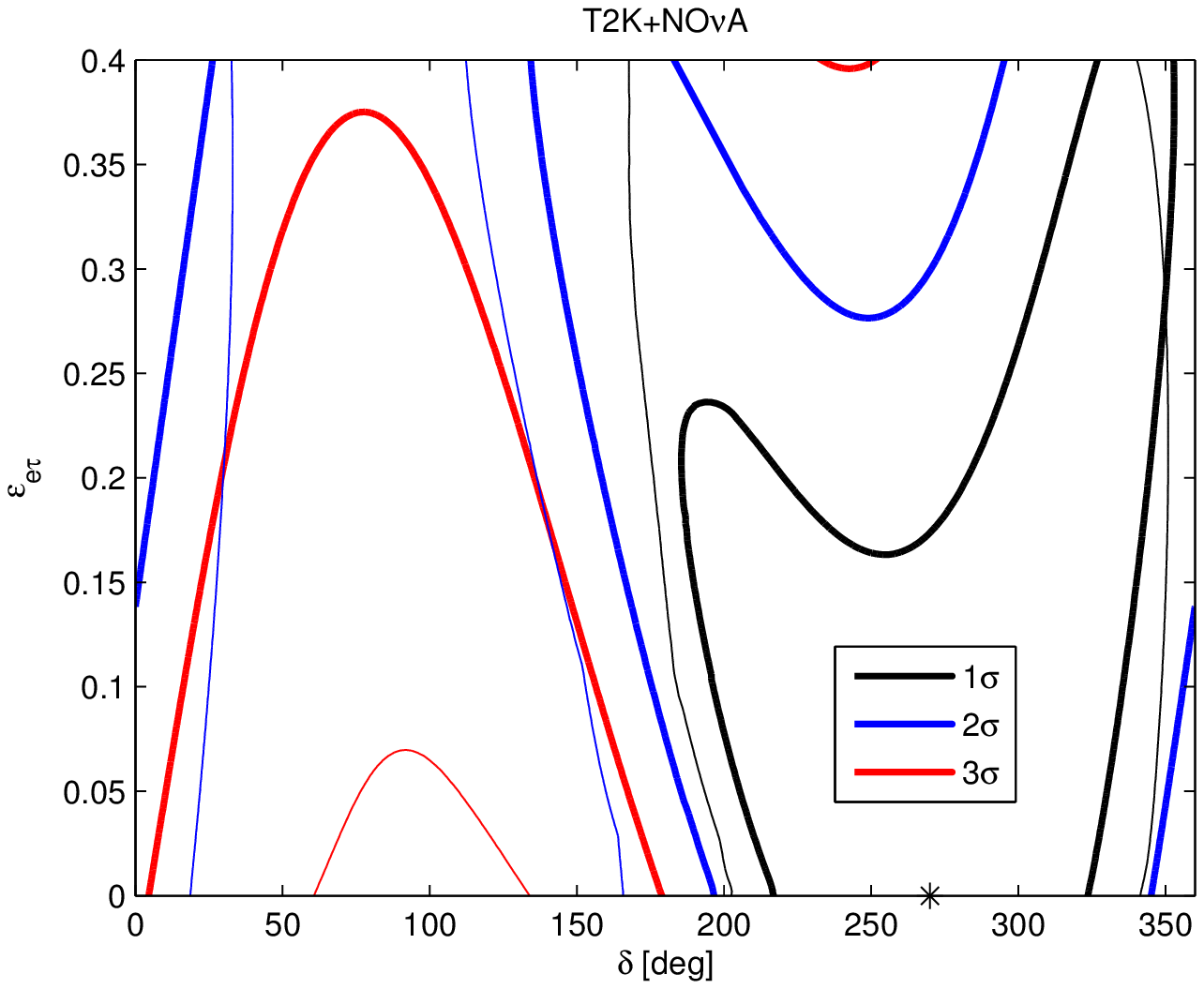}}
\subfigure[]{\includegraphics[width=0.49\textwidth]{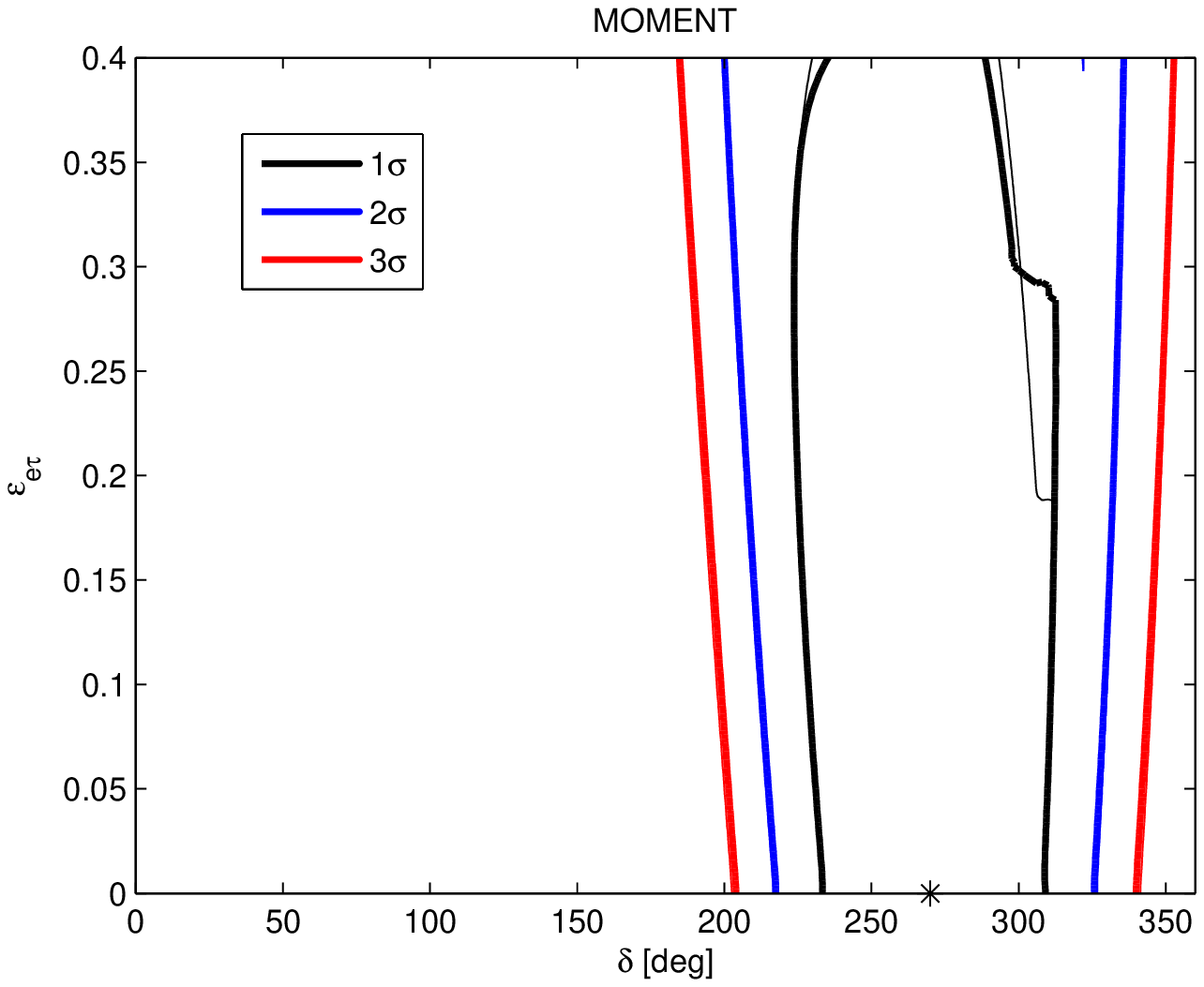}}
\subfigure[]{\includegraphics[width=0.49\textwidth]{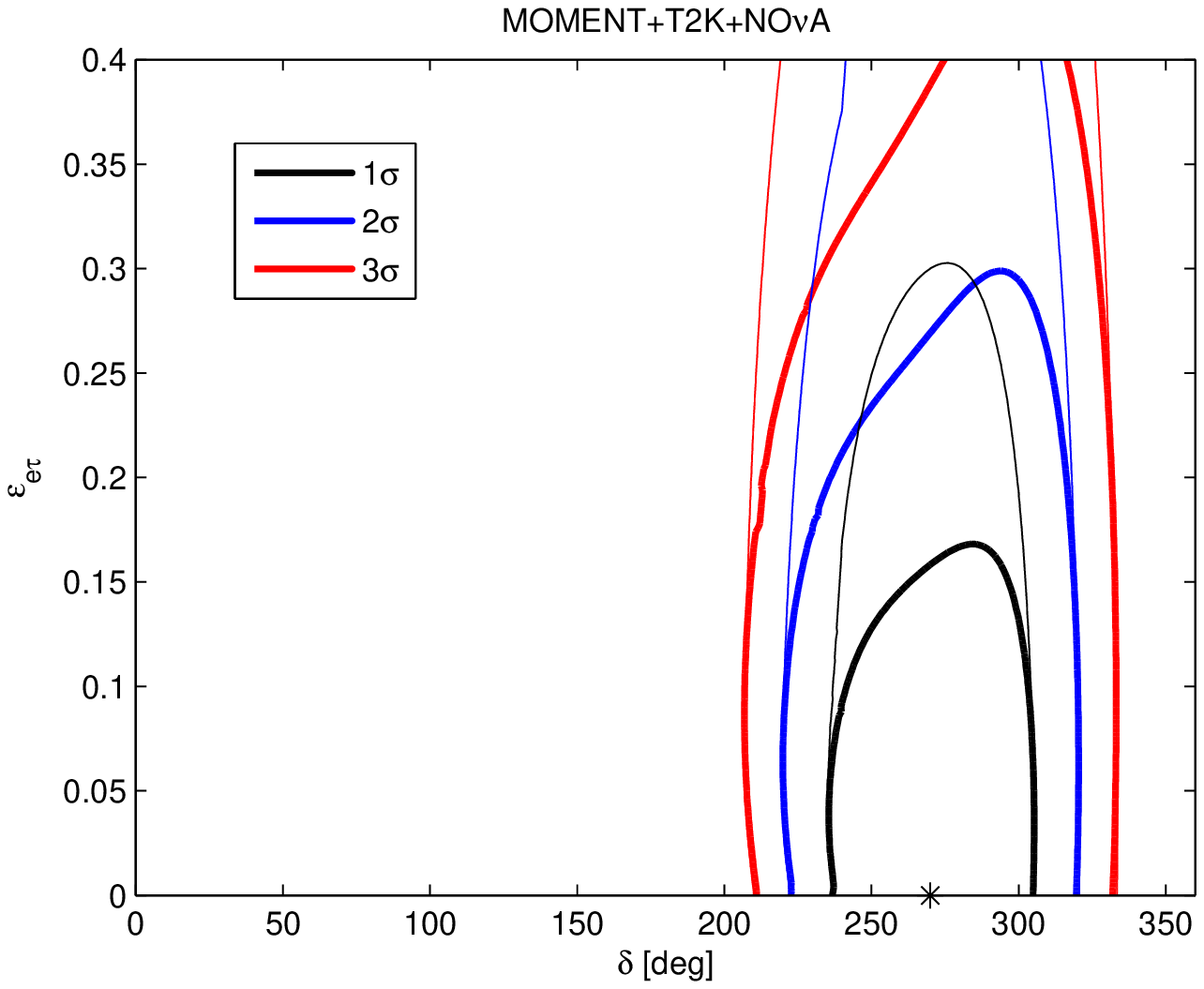}}
\end{center}
\vspace{2cm} \caption[]{Similar to Fig. \ref{ee} except that $\epsilon_{ee}$ is replaced by $|\epsilon_{e\tau}|$. The phase of $|\epsilon_{e\tau}|$ varies in $[0,2\pi]$. The $3 \sigma$ upper bound on $|\epsilon_{e\tau}|$ is 1.2 which lies outside the frames of these figures. {In drawing thick lines, all the $\epsilon_{\alpha \beta}$ except $\epsilon_{e\tau}$ are fixed to zero but when drawing the thin lines, we have allowed $\epsilon_{ee}$  to vary within the uncertainties shown in Eq. (\ref{boundsonE1}) using pull method.}}
\label{etau}
\end{figure}
Fig \ref{ee} shows the allowed region in $\delta$ and $\epsilon_{ee}$.  Neutrino mixing and mass splitting parameters are set to their best values shown in Ref. \cite{Gonzalez-Garcia:2014bfa} and their uncertainties (also taken from Ref. \cite{Gonzalez-Garcia:2014bfa}) are treated by the pull method. The ordering is taken to be normal and assumed to be known. Figs \ref{ee}-a, \ref{ee}-b and \ref{ee}-c respectively show the results from NO$\nu$A and T2K experiments, from the MOMENT experiment  and from the combined results. In drawing the thick lines, the rest of $\epsilon$ are fixed to zero.
Fig. \ref{ee}-b demonstrates that the MOMENT experiment is not very sensitive to $\epsilon_{ee}$ which helps to solve the degeneracy between $\delta$ and $\epsilon_{ee}$. From Fig. \ref{ee}-c, we observe that once we combine the MOMENT results with NO$\nu$A and T2K, CP-violation for $\delta=270^\circ$ can be established at  better than 3$\sigma$ C.L. even allowing nonzero $\epsilon_{ee}$.
Moreover, adding results of MOMENT, the $3\sigma$ bound on $\epsilon_{ee}$ slightly improves.
{As demonstrated in Fig 4 of \cite{Pilar-DUNE}, when $\epsilon_{e\tau}$ and $\epsilon_{ee}$ are simultaneously nonzero, a degeneracy appears that allows large values of $\epsilon_{ee}$ and $\epsilon_{e\tau}$ to hide from long baseline experiment results. To study this effect, we have superimposed the thin lines which are drawn applying pull method on $\epsilon_{e\tau}$ and allowing its phase to vary in $[0,2\pi]$. As expected the difference for MOMENT is small, but for NO$\nu$A+T2K the difference can be significant.
Fig \ref{ee}-c shows that when the NO$\nu$A+T2K results are combined with the MOMENT results  the determination of $\delta$ is not much affected but the uncertainty of $\epsilon_{ee}$ is increased by degeneracy between $\epsilon_{e\tau}$ and $\epsilon_{ee}$ that has been pointed out in Fig 4 of \cite{Pilar-DUNE}}

Fig. \ref{etau} is similar to Fig \ref{ee} except that it respectively shows the allowed ranges of  $\epsilon_{e\tau}-\delta$, allowing the phase of $\epsilon_{e\tau}$ to vary in $[0,2\pi]$. Drawing the thin lines, pull method is applied on $\epsilon_{ee}$ with 1$\sigma$ range $0<\epsilon_{ee}<0.93$ \cite{Maltoni}. Thick lines are drawn fixing $\epsilon_{ee}=0$. As expected turning on and off $\epsilon_{ee}$ does not make a significant difference for MOMENT
but T2K+NO$\nu$A results significantly change.  Comparing Figs \ref{etau}-a and \ref{etau}-c, we observe that when $\epsilon_{ee}$ is turned off, combining the MOMENT results with T2K+NO$\nu$A can help to significantly  improve the bound on $|\epsilon_{e\tau}|$. When $\epsilon_{ee}$ varies within its 1$\sigma$ C.L., determination of $\epsilon_{e\tau}$ worsens but still MOMENT can help to determine $\delta$ and rule out the wrong solution for $\delta$.
 \begin{figure}
\begin{center}
\subfigure[]{\includegraphics[width=0.49\textwidth]{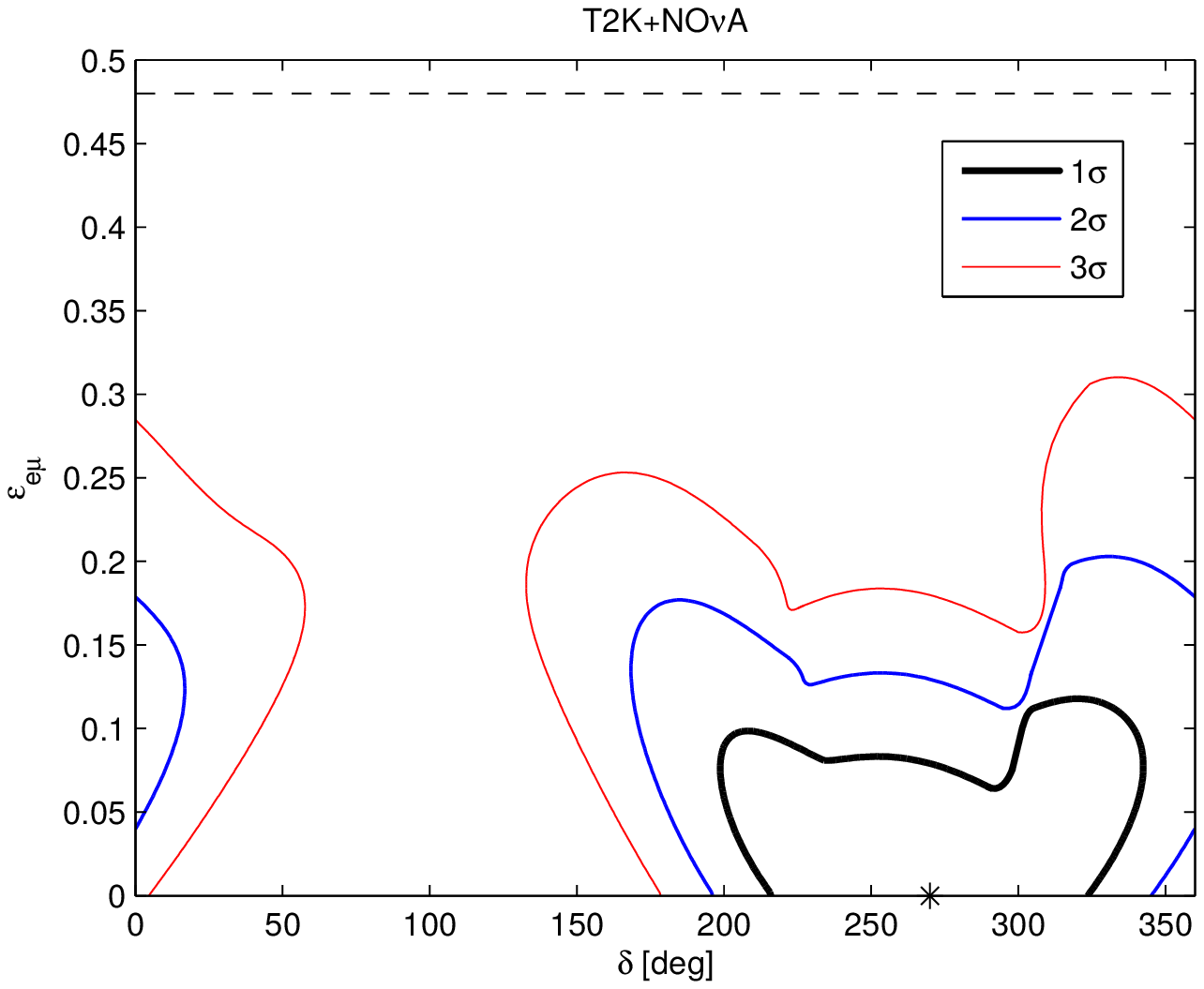}}
\subfigure[]{\includegraphics[width=0.49\textwidth]{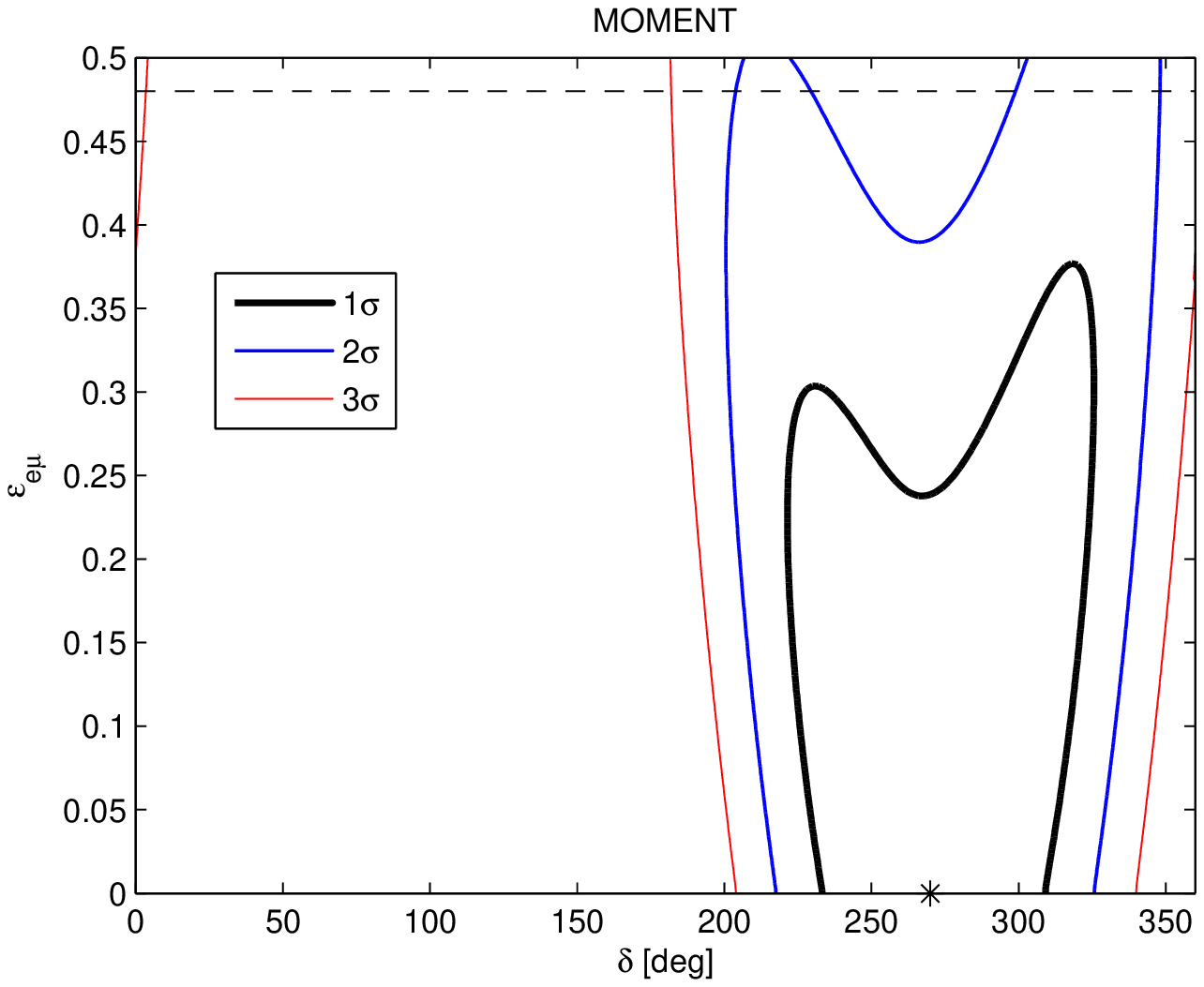}}
\subfigure[]{\includegraphics[width=0.49\textwidth]{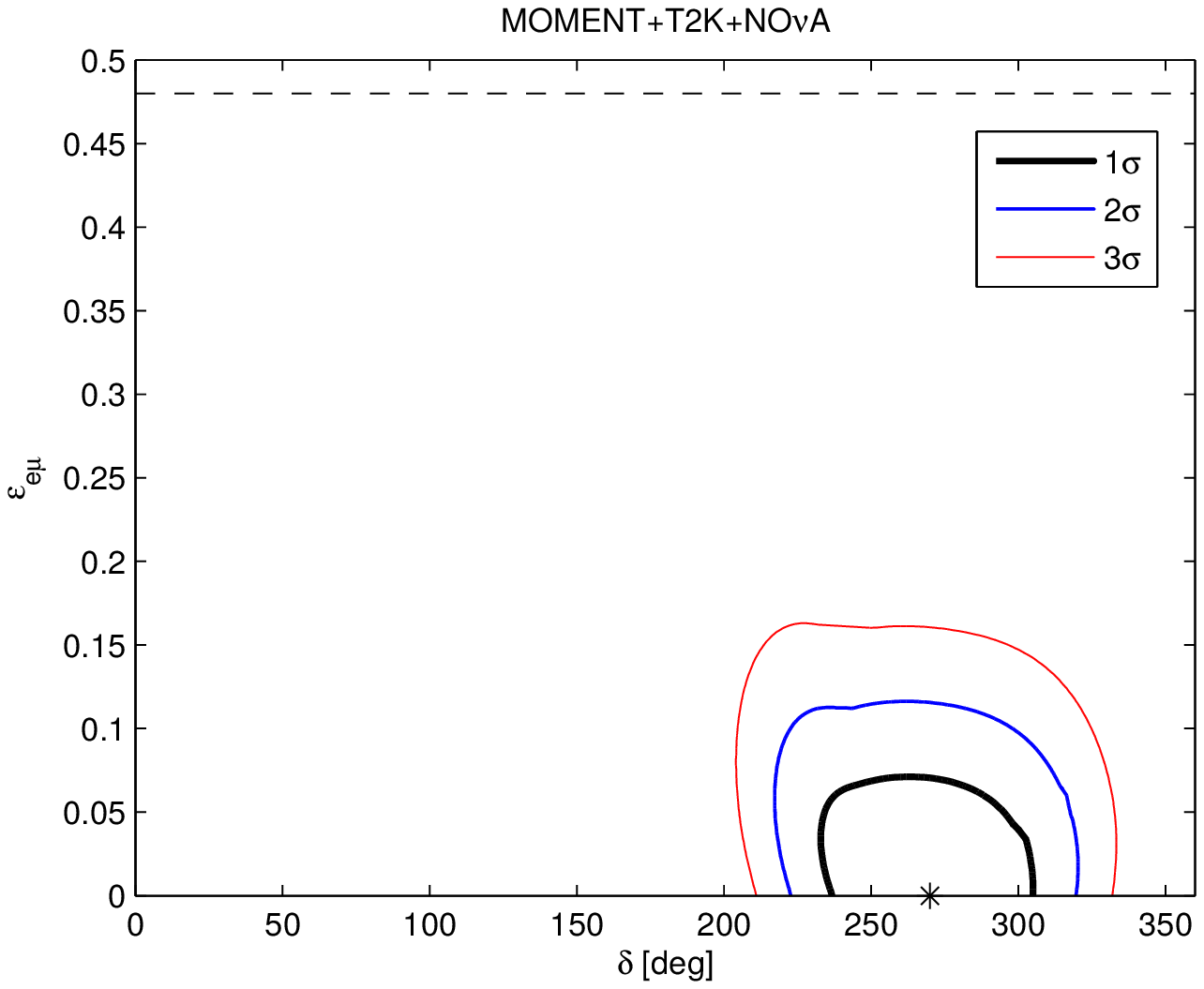}}
\end{center}
\vspace{2cm} \caption[]{Similar to Fig. \ref{ee} except that $\epsilon_{ee}$ is replaced by $|\epsilon_{e\mu}|$. The phase of $|\epsilon_{e\mu}|$ varies in $[0,2\pi]$.}
\label{emu}
\end{figure}

Figs \ref{emu} is similar to Figs \ref{ee}  and \ref{etau} except that it shows the allowed ranges of  $\epsilon_{e\mu}-\delta$, allowing the phase of
$\epsilon_{e\mu}$ to vary in $[0,2\pi]$. We have fixed all the rest of $\epsilon_{\alpha \beta}$ to zero.
Notice that combining the NO$\nu$A and T2K results with the results from MOMENT not only rules out the wrong solution for $\delta$ but also
improves the bound on $|\epsilon_{e\mu}|$.

In all above cases we have assumed normal mass ordering and have assumed that the mass ordering will be determined by other experiments such as JUNO. We repeated the analysis for inverted mass ordering and  found the same  overall results.
Ref. \cite{Pilar-Moment} show that MOMENT alone can determine the mass ordering. We found that this result is robust even when NSI are turned on and values of $\epsilon_{\alpha \beta}$ are  allowed to vary in the range displayed in Eqs. (\ref{boundsonE},\ref{boundsonE1}).
The wrong mass ordering can be ruled out at 95$\%$ C.L. by MOMENT alone.

%As shown in \cite{us}, the JUNO and RENO-50 experiments
%cannot tell the difference between LMA and LMA-Dark solutions once mass ordering is also flipped.
%Unlike those medium baseline reactor experiments, the MOMENT experiment is not much sensitive to the octant of $\theta_{12}$. Instead the MOMENT experiment is sensitive to
%$\epsilon_{\tau \tau}-\epsilon_{ee}\simeq \epsilon_{\mu \mu}-\epsilon_{ee}\sim 1$ required for the LMA-Dark solution of the solar neutrino anomaly. We have shown that MOMENT
%can rule out the solution with wrong mass ordering and  $\epsilon_{\tau \tau}-\epsilon_{ee}\simeq \epsilon_{\mu \mu}-\epsilon_{ee}=(5.5,7.0) ~[(4.7,9.0)]$, corresponding
 %1$\sigma$ [3$\sigma$] LMA-Dark solutions at 90\% C.L. [95 \% C.L.]. To obtain this result we have allowed all $\epsilon_{\alpha \beta}$ to vary within the range of LMA-Dark
 %solution shown in \cite{Maltoni}.

% \begin{figure}
%\begin{center}
%\subfigure[]{\includegraphics[width=0.49\textwidth]{fig6a}}
%\subfigure[]{\includegraphics[width=0.49\textwidth]{fig6b}}
%\end{center}
%\vspace{2cm} \caption[]{SF= 0.1 and 10 $\%$}

%\label{SF}

%\end{figure}
\section{Summary \label{sum}}

Long baseline neutrino experiments such as NO$\nu$A and DUNE are sensitive to matter effects. To extract the value of the Dirac CP-violating phase, $\delta$, the matter effects therefore have to be known and properly taken into account. Non-standard interaction of neutrinos with matter can induce degeneracies in determination of $\delta$. For example, at NO$\nu$A and T2K, the signatures of CP-violating scenario with $\delta=270^\circ$ within the SM ({\it i.e.,} $\epsilon_{\alpha \beta}=0$) can be mimicked by CP-conserving scenario ($\delta=0$ or $180^\circ$) with nonzero $\epsilon_{\alpha \beta}$. Even the upcoming state-of-the-art DUNE experiment cannot solve this degeneracy. We have studied how the proposed MOMENT experiment with $L=150$~km and $200~{\rm MeV}<E_\nu<600~{\rm MeV}$, which is also designed to extract $\delta$, can help to solve this degeneracy.  The results are shown in Figs. 1-6.

Because of relatively short baseline ($L\simeq 150$~km) and relatively low energy, the sensitivity of the MOMENT experiment to matter effects, either standard or non-standard, will be quite limited
($\sqrt{2} G_F N_e  L \ll 1$ and $\Delta m_{31}^2/E_\nu \gg \sqrt{2} G_F N_e $). Thus, MOMENT alone cannot put strong bounds on $\epsilon_{\alpha \beta}$. On the other, the low sensitivity to the matter effects means that, unlike at NO$\nu$A, turning on the NSI parameters at the MOMENT experiment cannot mimic the effects of CP-violating phase $\delta$ so the MOMENT experiment can help to solve the degeneracy. Comparing Fig 1 and 2, we observe that while in the presence of NSI, NO$\nu$A and T2K cannot determine $\delta$ and/or the octant of $\theta_{23}$, once the results of the MOMENT experiment are combined  with those of  T2K and NO$\nu$A, CP-violation can be established at better than 3$\sigma$ for $\delta=270^\circ$ and the octant of $\theta_{23}$ can be determined at $2\sigma$. These results are obtained by setting the true values of $\epsilon_{\alpha \beta}$ equal to zero, but treating their present uncertainties shown in Eqs (\ref{boundsonE},\ref{boundsonE1}) with pull method.
Fig. \ref{SF} shows the dependence of the performance of the MOMENT experiment on the background Suppression Factor (SF).  Determination of $\delta$, both with and without NSI, is not so much
sensitive to background SF and even with a modest suppression factor of 10 \%, $\delta$ can be determined. However to determine the octant of $\theta_{23}$ in the presence of NSI, SF should be better than 10 \%.

Although the  MOMENT experiment alone cannot give a significant bound on $|\epsilon_{\alpha \beta}|$, we have found that combining the MOMENT results with
NO$\nu$A and T2K can significantly improve the bounds on $|\epsilon_{e\mu}|$ and on $|\epsilon_{e\tau}|$. The present $3\sigma$ bound on $|\epsilon_{e\mu}|$ from the present global neutrino analysis is 0.48.
While T2K and NO$\nu$A can improve the 3$\sigma$ bound to 0.35, once combined with the MOMENT results the bound will be improved to 0.15. Setting the rest of elements of $\epsilon_{\alpha \beta}$ equal to zero, the combined bound  from  MOMENT, T2K and NO$\nu$A on $|\epsilon_{e\tau}|$ will be 0.45 which will be an improvement of factor 2.7 relative to the present $3\sigma$ bound from global analysis of neutrino oscillation data \cite{Maltoni}. The sensitivity to $\epsilon_{\mu \tau}$ and $\epsilon_{\mu \mu}-\epsilon_{\tau \tau}$ in all these three experiments is only mild.

\subsection*{Acknowledgments}
The authors would like to thank Dr Pilar Coloma for useful comments.
They acknowledge partial support from the  European Union FP7 ITN INVISIBLES (Marie Curie Actions, PITN- GA-2011- 289442). YF acknowledges ICTP, especially its  associate office, for generous  financial support and  the hospitality of its staff.

%%%%%%%%%%%%%%%%%%%%%%%%%%%%%%%%%%%%%%%%%%%%%%%%%%
%%%%%%%%%%%%%%%%%%%%%%%%%%%55KETABNAMEH%%%%%%%%%%%
%%%%%%%%%%%%%%%%%%%%%%%%%%%%%%%%%%%%%%%%%%%%%%%%%%


\begin{thebibliography}{99}

\bibitem{An:2015jdp}
  F.~An {\it et al.} [JUNO Collaboration],
  %``Neutrino Physics with JUNO,''
  J.\ Phys.\ G {\bf 43} (2016) 030401
  doi:10.1088/0954-3899/43/3/030401
  [arXiv:1507.05613 [physics.ins-det]].
  %%CITATION = doi:10.1088/0954-3899/43/3/030401;%%
  %23 citations counted in INSPIRE as of 07 Feb 2016

\bibitem{Kim:2014rfa}
  S.~B.~Kim,
  %``New results from RENO and prospects with RENO-50,''
  Nucl.\ Part.\ Phys.\ Proc.\  {\bf 265-266} (2015) 93
  doi:10.1016/j.nuclphysbps.2015.06.024
  [arXiv:1412.2199 [hep-ex]].
  %%CITATION = doi:10.1016/j.nuclphysbps.2015.06.024;%%
  %28 citations counted in INSPIRE as of 07 févr. 2016

\bibitem{Aartsen:2014oha}
  M.~G.~Aartsen {\it et al.} [IceCube PINGU Collaboration],
  %``Letter of Intent: The Precision IceCube Next Generation Upgrade (PINGU),''
  arXiv:1401.2046 [physics.ins-det].
  %%CITATION = ARXIV:1401.2046;%%
  %100 citations counted in INSPIRE as of 07 févr. 2016

\bibitem{INO}
S.~Ahmed {\it et al.} [ICAL Collaboration],
  %``Physics Potential of the ICAL detector at the India-based Neutrino Observatory (INO),''
  arXiv:1505.07380 [physics.ins-det].
  %%CITATION = ARXIV:1505.07380;%%
  \bibitem{Thomas}
    M.~C.~Gonzalez-Garcia, M.~Maltoni and T.~Schwetz,
  %``Updated fit to three neutrino mixing: status of leptonic CP violation,''
  JHEP {\bf 1411} (2014) 052
  doi:10.1007/JHEP11(2014)052
  [arXiv:1409.5439 [hep-ph]];
  %%CITATION = doi:10.1007/JHEP11(2014)052;%%
  J.~Bergstrom, M.~C.~Gonzalez-Garcia, M.~Maltoni and T.~Schwetz,
  %``Bayesian global analysis of neutrino oscillation data,''
  JHEP {\bf 1509} (2015) 200
  doi:10.1007/JHEP09(2015)200
  [arXiv:1507.04366 [hep-ph]];
  %%CITATION = doi:10.1007/JHEP09(2015)200;%%
   A.~Palazzo,
  %``3-flavor and 4-flavor implications of the latest T2K and NO$\nu$A electron (anti-)neutrino appearance results,''
  Phys.\ Lett.\ B {\bf 757} (2016) 142
  doi:10.1016/j.physletb.2016.03.061
  [arXiv:1509.03148 [hep-ph]].
  %%CITATION = doi:10.1016/j.physletb.2016.03.061;%%
  \bibitem{Forero:2014bxa}
  D.~V.~Forero, M.~Tortola and J.~W.~F.~Valle,
  %``Neutrino oscillations refitted,''
  Phys.\ Rev.\ D {\bf 90} (2014) 9,  093006
  doi:10.1103/PhysRevD.90.093006
  [arXiv:1405.7540 [hep-ph]].
  %%CITATION = doi:10.1103/PhysRevD.90.093006;%%
  %251 citations counted in INSPIRE as of 01 Mar 2016
  \bibitem{Lisi}
  F.~Capozzi, E.~Lisi, A.~Marrone, D.~Montanino and A.~Palazzo,
  %``Neutrino masses and mixings: Status of known and unknown $3\nu$ parameters,''
  arXiv:1601.07777 [hep-ph].
  %%CITATION = ARXIV:1601.07777;%%
\bibitem{DUNE}
R.~Acciarri {\it et al.} [DUNE Collaboration],
  %``Long-Baseline Neutrino Facility (LBNF) and Deep Underground Neutrino Experiment (DUNE) Conceptual Design Report Volume 2: The Physics Program for DUNE at LBNF,''
  arXiv:1512.06148 [physics.ins-det];
  %%CITATION = ARXIV:1512.06148;%%
  R.~Acciarri {\it et al.} [DUNE Collaboration],
  %``Long-Baseline Neutrino Facility (LBNF) and Deep Underground Neutrino Experiment (DUNE) Conceptual Design Report Volume 1: The LBNF and DUNE Projects,''
  arXiv:1601.05471 [physics.ins-det];
  %%CITATION = ARXIV:1601.05471;%%
  %1 citations counted in INSPIRE as of 07 Feb 2016
 J.~Strait {\it et al.} [DUNE Collaboration],
  %``Long-Baseline Neutrino Facility (LBNF) and Deep Underground Neutrino Experiment (DUNE) Conceptual Design Report Volume 3: Long-Baseline Neutrino Facility for DUNE June 24, 2015,''
  arXiv:1601.05823 [physics.ins-det];
  %%CITATION = ARXIV:1601.05823;%%
  R.~Acciarri {\it et al.} [DUNE Collaboration],
  %``Long-Baseline Neutrino Facility (LBNF) and Deep Underground Neutrino Experiment (DUNE) Conceptual Design Report, Volume 4 The DUNE Detectors at LBNF,''
  arXiv:1601.02984 [physics.ins-det].
  %%CITATION = ARXIV:1601.02984;%%
  \bibitem{T2HK}
   K.~Abe {\it et al.},
  %``Letter of Intent: The Hyper-Kamiokande Experiment --- Detector Design and Physics Potential ---,''
  arXiv:1109.3262 [hep-ex];
  %%CITATION = ARXIV:1109.3262;%%
  K.~Abe {\it et al.} [Hyper-Kamiokande Working Group Collaboration],
  %``A Long Baseline Neutrino Oscillation Experiment Using J-PARC Neutrino Beam and Hyper-Kamiokande,''
  arXiv:1412.4673 [physics.ins-det].
  %%CITATION = ARXIV:1412.4673;%%


\bibitem{Patterson:2015xja}
  R.~B.~Patterson,
  %``Prospects for Measurement of the Neutrino Mass Hierarchy,''
  Ann.\ Rev.\ Nucl.\ Part.\ Sci.\  {\bf 65} (2015) 177
  doi:10.1146/annurev-nucl-102014-021916
  [arXiv:1506.07917 [hep-ex]].
  %%CITATION = doi:10.1146/annurev-nucl-102014-021916;%%
\bibitem{Yas}
Y.~Farzan and A.~Y.~Smirnov,
  %``Leptonic unitarity triangle and CP violation,''
  Phys.\ Rev.\ D {\bf 65} (2002) 113001
  doi:10.1103/PhysRevD.65.113001
  [hep-ph/0201105].
  %%CITATION = doi:10.1103/PhysRevD.65.113001;%%
  %86 citations counted in INSPIRE as of 01 Mar 2016

  \bibitem{Palazzo:2011vg}
  A.~Palazzo,
  %``Hint of non-standard dynamics in solar neutrino conversion,''
  Phys.\ Rev.\ D {\bf 83} (2011) 101701
  doi:10.1103/PhysRevD.83.101701
  [arXiv:1101.3875 [hep-ph]].
  %%CITATION = doi:10.1103/PhysRevD.83.101701;%%
\bibitem{Ge}
J.~Evslin, S.~F.~Ge and K.~Hagiwara,
  %``The Leptonic CP Phase from T2(H)K and Muon Decay at Rest,''
  JHEP {\bf 1602} (2016) 137
  doi:10.1007/JHEP02(2016)137
  [arXiv:1506.05023 [hep-ph]].
  %%CITATION = doi:10.1007/JHEP02(2016)137;%%
  %2 citations counted in INSPIRE as of 01 Mar 2016
\bibitem{Cao:2014bea}
  J.~Cao {\it et al.},
  %``Muon-decay medium-baseline neutrino beam facility,''
  Phys.\ Rev.\ ST Accel.\ Beams {\bf 17} (2014) 090101
  doi:10.1103/PhysRevSTAB.17.090101
  [arXiv:1401.8125 [physics.acc-ph]].
  %%CITATION = doi:10.1103/PhysRevSTAB.17.090101;%%
  %8 citations counted in INSPIRE as of 04 Jan 2016


  %%CITATION = ARXIV:1601.07777;%%
\bibitem{Pilar-Moment}
  M.~Blennow, P.~Coloma and E.~Fern�ndez-Martinez,
  %``The MOMENT to search for CP violation,''
  arXiv:1511.02859 [hep-ph].
  %%CITATION = ARXIV:1511.02859;%%

\bibitem{Yasaman1}
Y.~Farzan,
  %``A model for large non-standard interactions of neutrinos leading to the LMA-Dark solution,''
  Phys.\ Lett.\ B {\bf 748} (2015) 311
  doi:10.1016/j.physletb.2015.07.015
  [arXiv:1505.06906 [hep-ph]].
  %%CITATION = doi:10.1016/j.physletb.2015.07.015;%%
\bibitem{Yasaman2}
Y.~Farzan and I.~M.~Shoemaker,
  %``Lepton Flavor Violating Non-Standard Interactions via Light Mediators,''
  arXiv:1512.09147 [hep-ph].
  %%CITATION = ARXIV:1512.09147;%%

\bibitem{Pilar-DUNE}
P.~Coloma,
  %``Non-Standard Interactions in propagation at the Deep Underground Neutrino Experiment,''
  arXiv:1511.06357 [hep-ph].
  %%CITATION = ARXIV:1511.06357;%%
\bibitem{Andre-DUNE}
 A.~de Gouv�a and K.~J.~Kelly,
  %``Non-standard Neutrino Interactions at DUNE,''
  arXiv:1511.05562 [hep-ph].
  %%CITATION = ARXIV:1511.05562;%%

\bibitem{Poonam}
M.~Masud, A.~Chatterjee and P.~Mehta,
  %``Probing CP violation signal at DUNE in presence of non-standard neutrino interactions,''
  arXiv:1510.08261 [hep-ph].
  %%CITATION = ARXIV:1510.08261;%%
\bibitem{Huber}

  D.~V.~Forero and P.~Huber,
  %``Hints for leptonic CP violation or New Physics?,''
  arXiv:1601.03736 [hep-ph].
  %%CITATION = ARXIV:1601.03736;%%
\bibitem{diagonal}
J.~Kopp,
  %``Efficient numerical diagonalization of hermitian 3 x 3 matrices,''
  Int.\ J.\ Mod.\ Phys.\ C {\bf 19} (2008) 523
  doi:10.1142/S0129183108012303
  [physics/0610206].
  %%CITATION = doi:10.1142/S0129183108012303;%%
\bibitem{Kopp:2007ne}
  J.~Kopp, M.~Lindner, T.~Ota and J.~Sato,
  %``Non-standard neutrino interactions in reactor and superbeam experiments,''
  Phys.\ Rev.\ D {\bf 77} (2008) 013007
  doi:10.1103/PhysRevD.77.013007
  [arXiv:0708.0152 [hep-ph]].
  %%CITATION = doi:10.1103/PhysRevD.77.013007;%%
  %109 citations counted in INSPIRE as of 04 Jan 2016
\bibitem{Danny}
J.~Liao, D.~Marfatia and K.~Whisnant,
  %``Degeneracies in long-baseline neutrino experiments from nonstandard interactions,''
  arXiv:1601.00927 [hep-ph];
  %%CITATION = ARXIV:1601.00927;%%
{\it see also,}  M.~Masud and P.~Mehta,
  %``Non-standard interactions spoiling the CP violation sensitivity at DUNE and other long baseline experiments,''
  arXiv:1603.01380 [hep-ph].
  %%CITATION = ARXIV:1603.01380;%%

  \bibitem{Maltoni}
  M.~C.~Gonzalez-Garcia and M.~Maltoni,
  %``Determination of matter potential from global analysis of neutrino oscillation data,''
  JHEP {\bf 1309} (2013) 152
  doi:10.1007/JHEP09(2013)152
  [arXiv:1307.3092].
  %%CITATION = doi:10.1007/JHEP09(2013)152;%%
  %16 citations counted in INSPIRE as of 04 Jan 2016

  \bibitem{Mariam}
  O.~G.~Miranda, M.~A.~Tortola and J.~W.~F.~Valle,
  %``Are solar neutrino oscillations robust?,''
  JHEP {\bf 0610} (2006) 008
  doi:10.1088/1126-6708/2006/10/008
  [hep-ph/0406280];
  %%CITATION = doi:10.1088/1126-6708/2006/10/008;%%
  %124 citations counted in INSPIRE as of 01 Mar 2016
   F.~J.~Escrihuela, O.~G.~Miranda, M.~A.~Tortola and J.~W.~F.~Valle,
  %``Constraining nonstandard neutrino-quark interactions with solar, reactor and accelerator data,''
  Phys.\ Rev.\ D {\bf 80} (2009) 105009
   [Phys.\ Rev.\ D {\bf 80} (2009) 129908]
  doi:10.1103/PhysRevD.80.129908, 10.1103/PhysRevD.80.105009
  [arXiv:0907.2630 [hep-ph]].
  %%CITATION = doi:10.1103/PhysRevD.80.129908, 10.1103/PhysRevD.80.105009;%%
  %35 citations counted in INSPIRE as of 01 Mar 2016

    \bibitem{Fogli}
  G.~L.~Fogli, E.~Lisi, A.~Marrone, D.~Montanino and A.~Palazzo,
  %``Getting the most from the statistical analysis of solar neutrino oscillations,''
  Phys.\ Rev.\ D {\bf 66} (2002) 053010
  doi:10.1103/PhysRevD.66.053010
  [hep-ph/0206162].
  %%CITATION = doi:10.1103/PhysRevD.66.053010;%%

  \bibitem{us}
  P.~Bakhti and Y.~Farzan,
  %``Shedding light on LMA-Dark solar neutrino solution by medium baseline reactor experiments: JUNO and RENO-50,''
  JHEP {\bf 1407} (2014) 064
  doi:10.1007/JHEP07(2014)064
  [arXiv:1403.0744 [hep-ph]].
  %%CITATION = doi:10.1007/JHEP07(2014)064;%%
  \bibitem{wang}
Y. Wang. 2015. plenary talk at Invisibles 2015, Madrid, Spain, June 24.
    \bibitem{CI}
  P.~Huber and T.~Schwetz,
  %``A Low energy neutrino factory with non-magnetic detectors,''
  Phys.\ Lett.\ B {\bf 669} (2008) 294
  doi:10.1016/j.physletb.2008.10.009
  [arXiv:0805.2019 [hep-ph]].
  %%CITATION = doi:10.1016/j.physletb.2008.10.009;%%
 \bibitem{BurguetCastell:2005pa}
  J.~Burguet-Castell, D.~Casper, E.~Couce, J.~J.~Gomez-Cadenas and P.~Hernandez,
  %``Optimal beta-beam at the CERN-SPS,''
  Nucl.\ Phys.\ B {\bf 725} (2005) 306
  doi:10.1016/j.nuclphysb.2005.06.037
  [hep-ph/0503021].
  %%CITATION = doi:10.1016/j.nuclphysb.2005.06.037;%%
  %139 citations counted in INSPIRE as of 07 févr. 2016





 \bibitem{Paschos:2001np}
  E.~A.~Paschos and J.~Y.~Yu,
  %``Neutrino interactions in oscillation experiments,''
  Phys.\ Rev.\ D {\bf 65} (2002) 033002
  doi:10.1103/PhysRevD.65.033002
  [hep-ph/0107261].
  %%CITATION = doi:10.1103/PhysRevD.65.033002;%%
  %117 citations counted in INSPIRE as of 07 févr. 2016

\bibitem{Messier:1999kj}
  M.~D.~Messier,
  %``Evidence for neutrino mass from observations of atmospheric neutrinos with Super-Kamiokande,''
  UMI-99-23965.
  %%CITATION = UMI-99-23965;%%
  %20 citations counted in INSPIRE as of 07 Feb 2016

  \bibitem{MEMphys}
  L.~Agostino {\it et al.} [MEMPHYS Collaboration],
  %``Study of the performance of a large scale water-Cherenkov detector (MEMPHYS),''
  JCAP {\bf 1301} (2013) 024
  doi:10.1088/1475-7516/2013/01/024
  [arXiv:1206.6665 [hep-ex]].
  %%CITATION = doi:10.1088/1475-7516/2013/01/024;%%
 \bibitem{Itow:2001ee}
  Y.~Itow {\it et al.} [T2K Collaboration],
  %``The JHF-Kamioka neutrino project,''
  hep-ex/0106019.
  %%CITATION = HEP-EX/0106019;%%
  %937 citations counted in INSPIRE as of 05 Feb 2016
  \bibitem{Huber:2002mx}
  P.~Huber, M.~Lindner and W.~Winter,
  %``Superbeams versus neutrino factories,''
  Nucl.\ Phys.\ B {\bf 645}, 3 (2002)
  doi:10.1016/S0550-3213(02)00825-8
  [hep-ph/0204352].
  %%CITATION = doi:10.1016/S0550-3213(02)00825-8;%%
  %302 citations counted in INSPIRE as of 05 Feb 2016



\bibitem{Ishitsuka:2005qi}
  M.~Ishitsuka, T.~Kajita, H.~Minakata and H.~Nunokawa,
  %``Resolving neutrino mass hierarchy and CP degeneracy by two identical detectors with different baselines,''
  Phys.\ Rev.\ D {\bf 72} (2005) 033003
  doi:10.1103/PhysRevD.72.033003
  [hep-ph/0504026].
  %%CITATION = doi:10.1103/PhysRevD.72.033003;%%
  %171 citations counted in INSPIRE as of 05 Feb 2016

\bibitem{Ayres:2004js}
  D.~S.~Ayres {\it et al.} [NOvA Collaboration],
  %``NOvA: Proposal to build a 30 kiloton off-axis detector to study nu(mu) ---> nu(e) oscillations in the NuMI beamline,''
  hep-ex/0503053.
  %%CITATION = HEP-EX/0503053;%%
  %587 citations counted in INSPIRE as of 05 Feb 2016

\bibitem{Yang_2004}
	T. Yang and S. Woijcicki. 2004. Off-Axis-Note-SIM-30.
%\cite{Huber:2007ji}





\bibitem{Huber:2004ka}
  P.~Huber, M.~Lindner and W.~Winter,
  %``Simulation of long-baseline neutrino oscillation experiments with GLoBES (General Long Baseline Experiment Simulator),''
  Comput.\ Phys.\ Commun.\  {\bf 167} (2005) 195
  doi:10.1016/j.cpc.2005.01.003
  [hep-ph/0407333].
  %%CITATION = doi:10.1016/j.cpc.2005.01.003;%%
  %326 citations counted in INSPIRE as of 04 Jan 2016


\bibitem{Huber:2007ji}
  P.~Huber, J.~Kopp, M.~Lindner, M.~Rolinec and W.~Winter,
  %``New features in the simulation of neutrino oscillation experiments with GLoBES 3.0: General Long Baseline Experiment Simulator,''
  Comput.\ Phys.\ Commun.\  {\bf 177} (2007) 432
  doi:10.1016/j.cpc.2007.05.004
  [hep-ph/0701187].
  %%CITATION = doi:10.1016/j.cpc.2007.05.004;%%
  %264 citations counted in INSPIRE as of 04 Jan 2016

\bibitem{Kopp:2006wp}
  J.~Kopp,
  %``Efficient numerical diagonalization of hermitian 3 x 3 matrices,''
  Int.\ J.\ Mod.\ Phys.\ C {\bf 19} (2008) 523
  doi:10.1142/S0129183108012303
  [physics/0610206].
  %%CITATION = doi:10.1142/S0129183108012303;%%
  %36 citations counted in INSPIRE as of 28 May 2016

  \bibitem{PREM}
  A. M. Dziewonski and D. L. Anderson, Phys. Earth Planet. Interiors \textbf{25} (1981) 297;
  R.~J.~Geller and T.~Hara,
  %``Geophysical aspects of very long baseline neutrino experiments,''
  Nucl.\ Instrum.\ Meth.\ A {\bf 503} (2003) 187
  doi:10.1016/S0168-9002(03)00670-3
  [hep-ph/0111342].
  %%CITATION = doi:10.1016/S0168-9002(03)00670-3;%%


\bibitem{Gonzalez-Garcia:2014bfa}
  M.~C.~Gonzalez-Garcia, M.~Maltoni and T.~Schwetz,
  %``Updated fit to three neutrino mixing: status of leptonic CP violation,''
  JHEP {\bf 1411} (2014) 052
  doi:10.1007/JHEP11(2014)052
  [arXiv:1409.5439 [hep-ph]];
  %%CITATION = doi:10.1007/JHEP11(2014)052;%%
  %250 citations counted in INSPIRE as of 05 f�vr. 2016
{\it see also,} www.nu-fit.org.

  \end{thebibliography}
\end{document}